\documentclass[10pt,a4paper]{article}

\usepackage{calc}
\usepackage[T1]{fontenc}
\usepackage[utf8]{inputenc}
\usepackage{amssymb,amsmath,amsthm,amsfonts} 
\usepackage{mathrsfs}  
\usepackage{bm}                                 
\usepackage{mathtools}                          
\usepackage{stmaryrd}                           
\usepackage{breqn}                              
\usepackage[textfont=footnotesize,labelfont={footnotesize,bf}]{caption}
\usepackage{subcaption}
\usepackage[title]{appendix}
\usepackage{booktabs}
\usepackage[inline]{enumitem}
\usepackage{graphicx}
\usepackage{graphbox}
\usepackage[dvipsnames]{xcolor}
\usepackage[top=30mm,right=30mm,left=30mm,bottom=30mm]{geometry}
\usepackage[colorlinks=true,linkcolor=blue,urlcolor=blue]{hyperref}
\usepackage[affil-it,auth-sc]{authblk}
\usepackage[colorinlistoftodos,textwidth=25mm,textsize=footnotesize]{todonotes}
\usepackage{lipsum}                             
\usepackage[section]{placeins}
\usepackage{tikz}
\usetikzlibrary{positioning}
\usepackage{scalefnt}
\usepackage[
	backend=biber,
	style=numeric,
	citestyle=numeric-comp,
	maxbibnames=99,
	minbibnames=99,
	maxcitenames=2,
	mincitenames=1,
	giveninits=true,
	sorting=none,
	sortlocale=auto,
	natbib=true,
	url=false,
	isbn=false,
	doi=true,
	eprint=true
]{biblatex}
\DeclareMathOperator{\sgn}{sgn}

\graphicspath{{./}{./figures/}}                 
\title{Bifurcations of an elastic disc coated with an elastic inextensible rod}

\author[1]{M. Gaibotti}
\author[2]{S.G.S.G. Mogilevskaya}
\author[1]{A. Piccolroaz}
\author[1]{D. Bigoni\footnote{Corresponding author: e-mail: \href{mailto:name@unitn.it}{bigoni@unitn.it}; phone: +39\,0461\,282507.}}

\affil[1]{Department of Civil, Environmental, and Mechanical Engineering, University of Trento, Italy}
\affil[2]{Department of Civil, Environmental and Geo-Engineering, University of Minnesota, 500 Pillsbury Drive S.E. Minneapolis, MN 55455-0116, USA.}

\date{}

\begin{document}

\maketitle

\begin{abstract}
An analytical solution is derived for the bifurcations
of an elastic disc that is constrained on the boundary with an isoperimetric Cosserat coating. The latter is treated as an elastic circular rod,
either perfectly or partially bonded (with a slip interface in the latter case) and is subjected to three different types of uniformly distributed radial loads (including hydrostatic pressure). The proposed solution technique employs complex potentials to
treat the disc’s interior and incremental Lagrangian
equations to describe the prestressed elastic rod
modelling the coating. The bifurcations of the disc occur with modes characterized by different circumferential wavenumbers, ranging between
ovalization and high-order waviness, as a function of the ratio between the elastic stiffness of the disc and the bending stiffness of its coating. The presented results find applications in various fields, such as coated fibres, mechanical rollers, and the growth and morphogenesis of plants and fruits.
\end{abstract}
\paragraph{Keywords}
Bifurcation \textperiodcentered\ 
Complex potentials \textperiodcentered\
Elastic rod theory \textperiodcentered

\section{Introduction}
\label{intro}
\begin{figure}[ht!]
\centering
\includegraphics[keepaspectratio, scale=0.033]{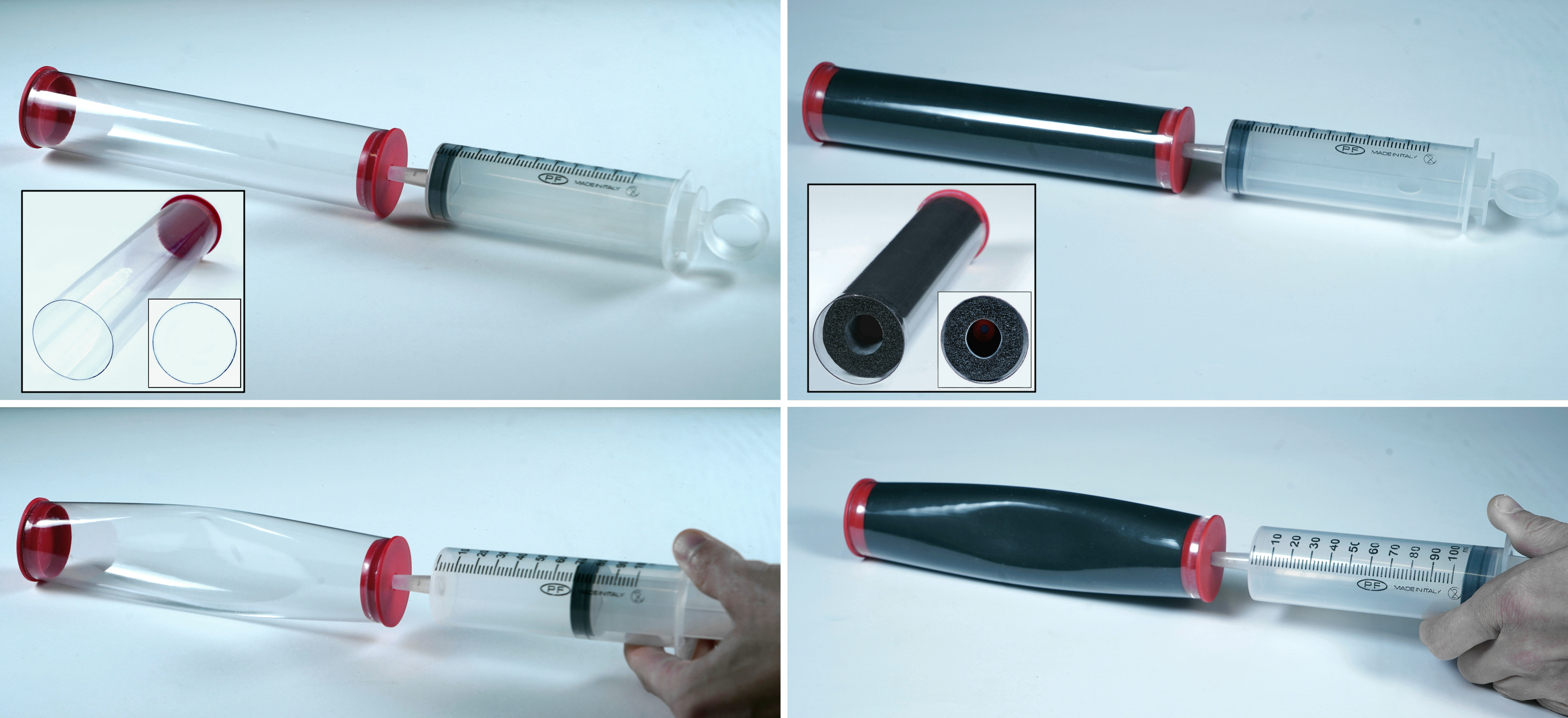}
\caption{A teaching model illustrates buckling in cylindrical shells. Left: a thin shell (undeformed in the upper part, 42\,mm in diameter, 0.5\,mm wall thickness made up of plastic) buckles under external pressure when air is extracted from it (lower part), resulting in ovalization. Right: buckling is also induced in a cylindrical shell with an inner core (a tubular rubber foam); in this experiment, the core’s stiffness is insufficient to induce bifurcation in a small-wavelength mode, so that in both cases ovalization occurs. Note the position of the plunger inside the syringe showing that the pressure load is higher in the experiment with the internal core. The simple experimental set-up shown on the right highlights the idea beyond the mechanical modelling, in which a shell becomes a coating enclosing an elastic core.}
\label{peretta}
\end{figure}

When an elastic cylindrical shell or a circular rod is subject to a radial external pressure of sufficient intensity, buckling occurs via ovalization, as shown with a teaching model in figure ~\ref{peretta} on the left (where a segment of an acrylic polymer tube with a diameter of 42\,mm and a wall thickness of 0.5\,mm is used, maintaining a thickness-to-diameter ratio equivalent to that of a chicken eggshell). 
This buckling phenomenon is well known and has been analysed by
various researchers, among others \cite{timoshenko1970theory, stevens1952stability, biezeno1945generalized, boresi1955refinement}. In particular, a distinction has been introduced between three different mechanical models for the external uniform radial load \cite{bodner1958conservativeness,armenakas1963vibrations}:
\begin{itemize}
    \item[(i.)] \lq Hydrostatic' or \lq pressure' load, which always remains orthogonal to the structural element to which it is applied in any configuration (undeformed or deformed);
    \item[(ii.)] \lq Centrally directed' load, which acts on the structural element remaining always directed towards the initial centre of the ring;
    \item[(iii.)] \lq Constant directional' or, better, \lq dead' load which remains aligned parallel to the unit normal to the structural element to which it is applied in its undeformed configuration. 
\end{itemize}
All three above loads are conservative \cite{bodner1958conservativeness} and the difference between them emerges in the incremental equations, holding for departures from the trivial configuration, so that they lead
to remarkably different bifurcation loads. 

Bifurcation also occurs in the case when an elastically deformable core is present inside of
the shell (or the circular rod), as shown with a teaching model in figure ~\ref{peretta} on the right (where the core is a rubber foam used for pipe insulation). The simple experimental set-up inspires the
mechanical modelling that will be adopted in the present article, where the coating of the elastic core is provided by an axially inextensible elastic rod. This model is already well developed in mechanics \cite{metrikine1997three,hashin2002thin,rubin2021nonlinear}, including cases involving circular and elliptic geometries \cite{reissner1949reinforced,radok1955problems,babaev1966physically,nemish1966reinforced,shul1967stress,plakhotnyi1967stresses,dhir1968stresses,chernyshenko1968nonlinear,guz1969stress,tsurpal1974problems}, and has also been used to analyse bifurcation, but only with respect to a half space \cite{shield1994buckling,tarasovs2008buckling}, while bifurcation
for circular configurations has never been analysed.

The inner core inside the circular rod not only increases the buckling load (note the position of the plunger inside the syringe, which shows that the pressure for buckling is higher when the inner core is present), but also results in a complex bifurcation behaviour that allows for the emergence of short-wavelength wave modes, although this does not take place in the case shown in figure ~\ref{peretta}, due to insufficient stiffness of the core.
The bifurcation problem of a coated elastic disc has been scarcely analysed and always only for hydrostatic pressure loading (i): in \cite{seide1962stability, seide1962buckling} (where a numerical, not analytical, solution is only found), under the assumption that the coating,
modelled as an elastic rod, cannot transmit shear stress to the inner disc (the imperfect bonding condition also considered by us); in \cite{herrmann1965buckling}, where the coating is modelled as an elastic shell, either
fully or partially bonded to the disc, and the obtained solution is analytical, though based on a number of mathematical simplifications. 
Therefore, the bifurcation of coated discs under radial
forces is still an open and almost unexplored problem.

\begin{figure}[ht!]
\centering
\includegraphics[keepaspectratio, scale=0.045]{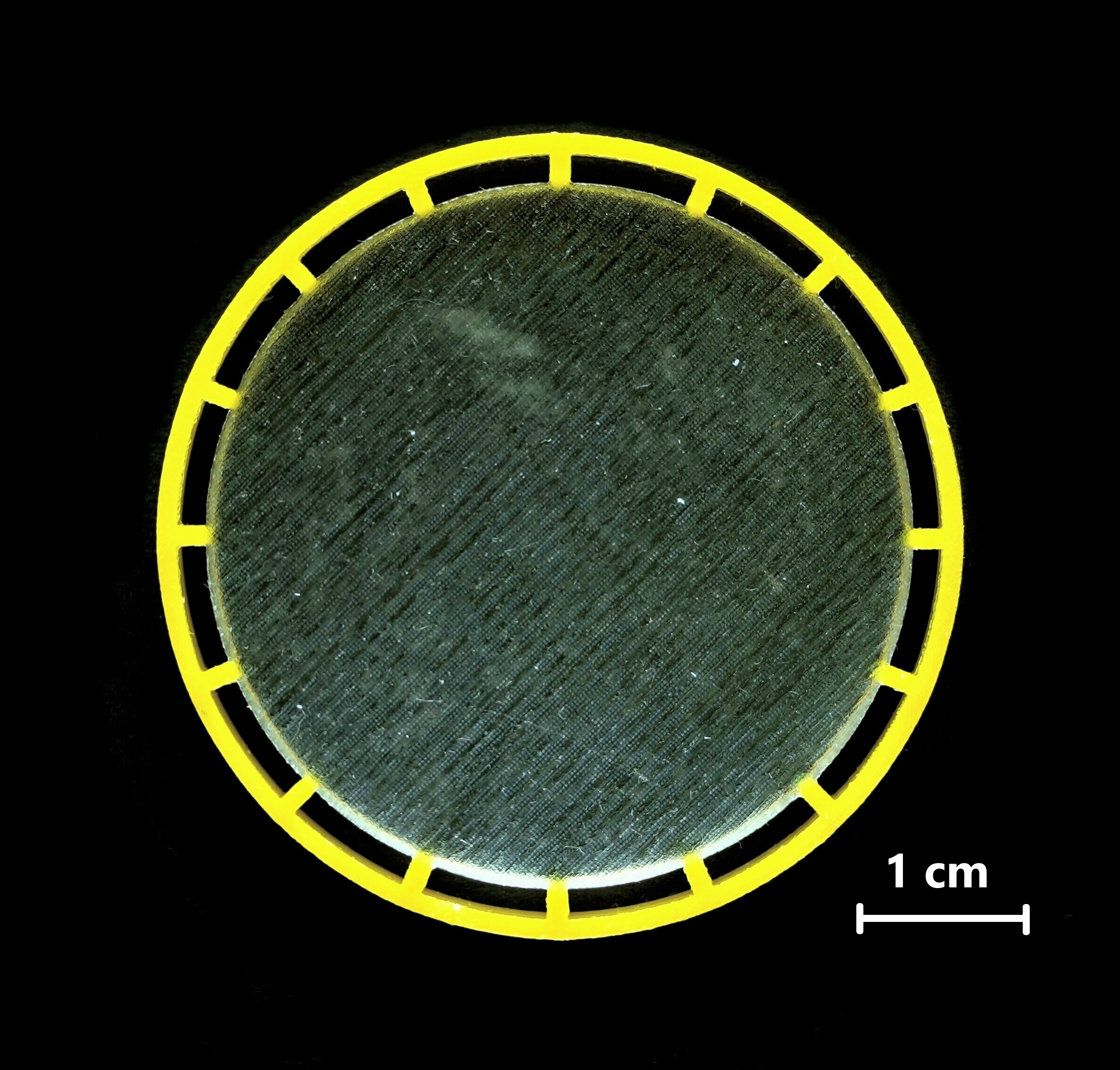}
\caption{A circular rubber disc (Agilus30\textsuperscript{\texttrademark}, 44\,mm in diameter) is  connected to a 
stiffer photopolymer (VeroYellow\textsuperscript{\texttrademark}, 1\,mm thick) coating through radial ligaments. The latter can transmit axial but only negligible transverse forces. Therefore, a model
of slip interface can be adopted to model the disc (manufactured with a Stratasys J750 3D printer) when radially loaded on its external surface. The piece is bioinspired by the joining through ligaments of the brain to the cranial vault.}
\label{pezzo_plastica}
\end{figure}

The present article employs the coating model for an elastic disc introduced in \cite{gaibotti2022isoDisk}, which can
be considered a specific case of the shell-coating model formulated in \cite{benveniste2001imperfect}. This model assumes
that the elastic disc is coated with an elastic rod that is axially inextensible and unshearable, thereby introducing a Cosserat and isoperimetric constraint for the disc. The analysis of buckling is carried out with all the three load variants (i)–(iii). Prior to bifurcation, the rod exhibits trivial equilibrium, characterized by a pure axial internal force. As a result, the elastic core, assumed to be
isotropic, remains unloaded before bifurcation and follows incremental equations obeying linear isotropic elasticity, characterized by Lamé constants $\lambda^{\text{d}}$ and $\mu^{\text{d}}$.  
Two transmission conditions are
analysed for the bonding between the elastic rod and the inner core, namely, perfect bonding,
where full continuity of radial and tangential displacements is enforced, and tangential slip
contact, in which the radial displacement is transmitted but the tangential is not, so that the shear stress at contact remains null. The latter condition can capture the behaviour of a partially detached coating. It can also model a coating attached at discrete points to the disc, as is the case of the piece (manufactured with a Stratasys J750 3D printer, following the multi-material Polyjet technique process, with a layer’s printing resolution of 27 $\mu$m) documented in the photo reported
in figure \ref{pezzo_plastica}. 

The mechanical model adopted in the present article allows for an analytical solution to the
bifurcation problem in a simple closed form, using Kolosov–Muskhelishvili complex potentials
for the core and incremental Lagrangian equations for the coating. The superiority of the
complex formalism becomes evident by comparing our closed-form and exact solution with
the approximate solution provided in \cite{herrmann1965buckling}.
The analysis demonstrates that imperfect bonding decreases the bifurcation threshold and that the hydrostatic-pressure model (i) results in the highest critical loads, while the centrally directed load model (ii) leads to the smallest. Significant differences in the bifurcation conditions discovered here highlight the importance of the role played by mechanical modelling of applied loads for accurate representations of load transfer mechanisms under structural deformation, a topic often underestimated and given here a new evidence. 

Results of the bifurcation analysis show that, in cases where the inner core is sufficiently compliant, the critical bifurcation mode corresponds to ovalization. However, as the ratio between
the stiffness of the elastic core and that of the coating rod increases, the modes display all possible waviness until they approach the vanishing-wavelength condition in the limit, where the stiffness
ratio approaches infinity.

\begin{figure}[ht!]
\centering
		\begin{tikzpicture}
		\node[inner sep=0pt] (figa2) at (0,0)
{\includegraphics[scale=0.1]{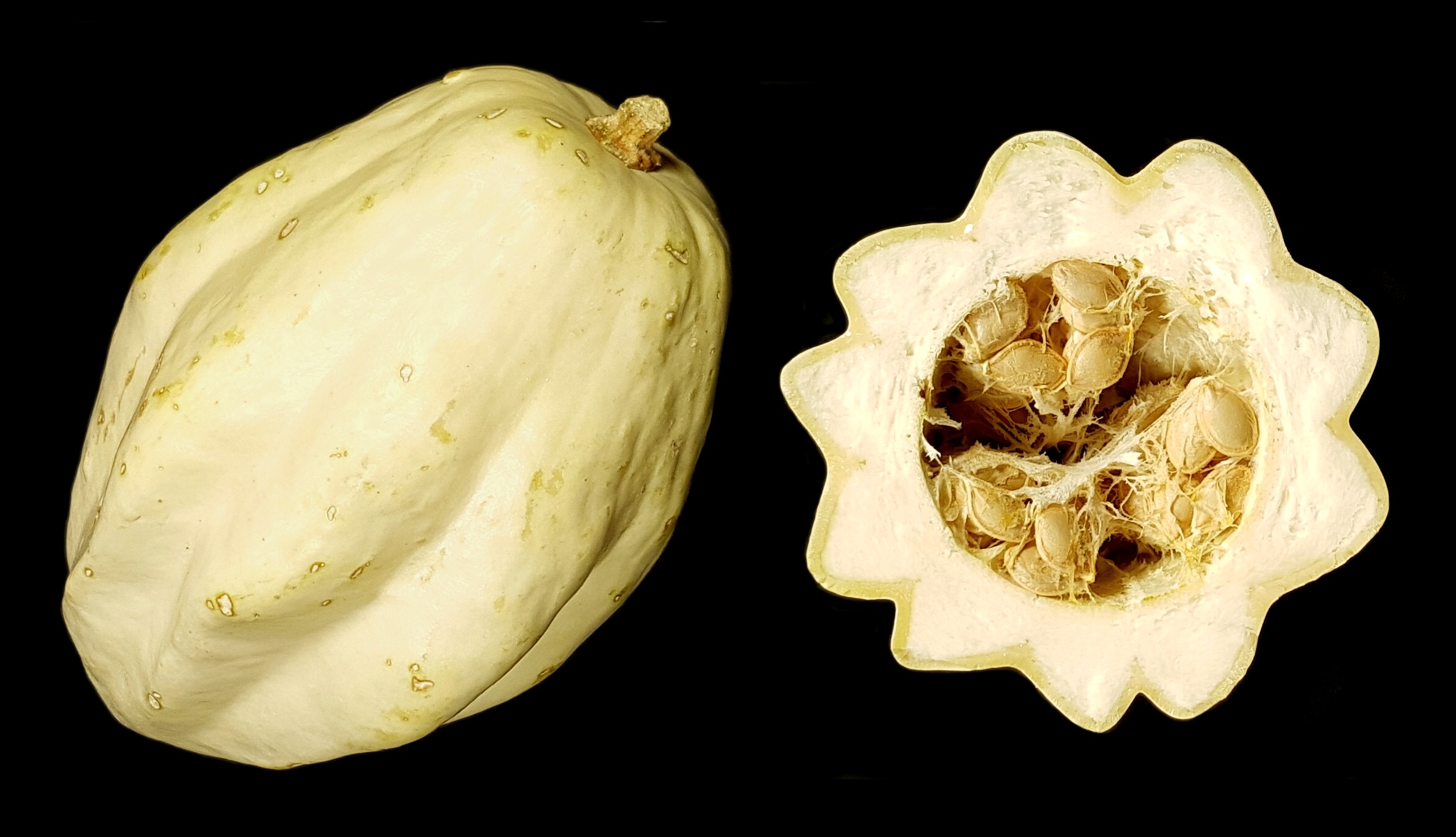}};
		\node[inner sep=1pt,fill=white,draw,rounded corners=1pt,below right=2mm and 2mm] at (figa2.north west) {(a)};
		\end{tikzpicture}
    \begin{tikzpicture}
		\node[inner sep=0pt] (figa2) at (0,0){\includegraphics[scale=0.17]{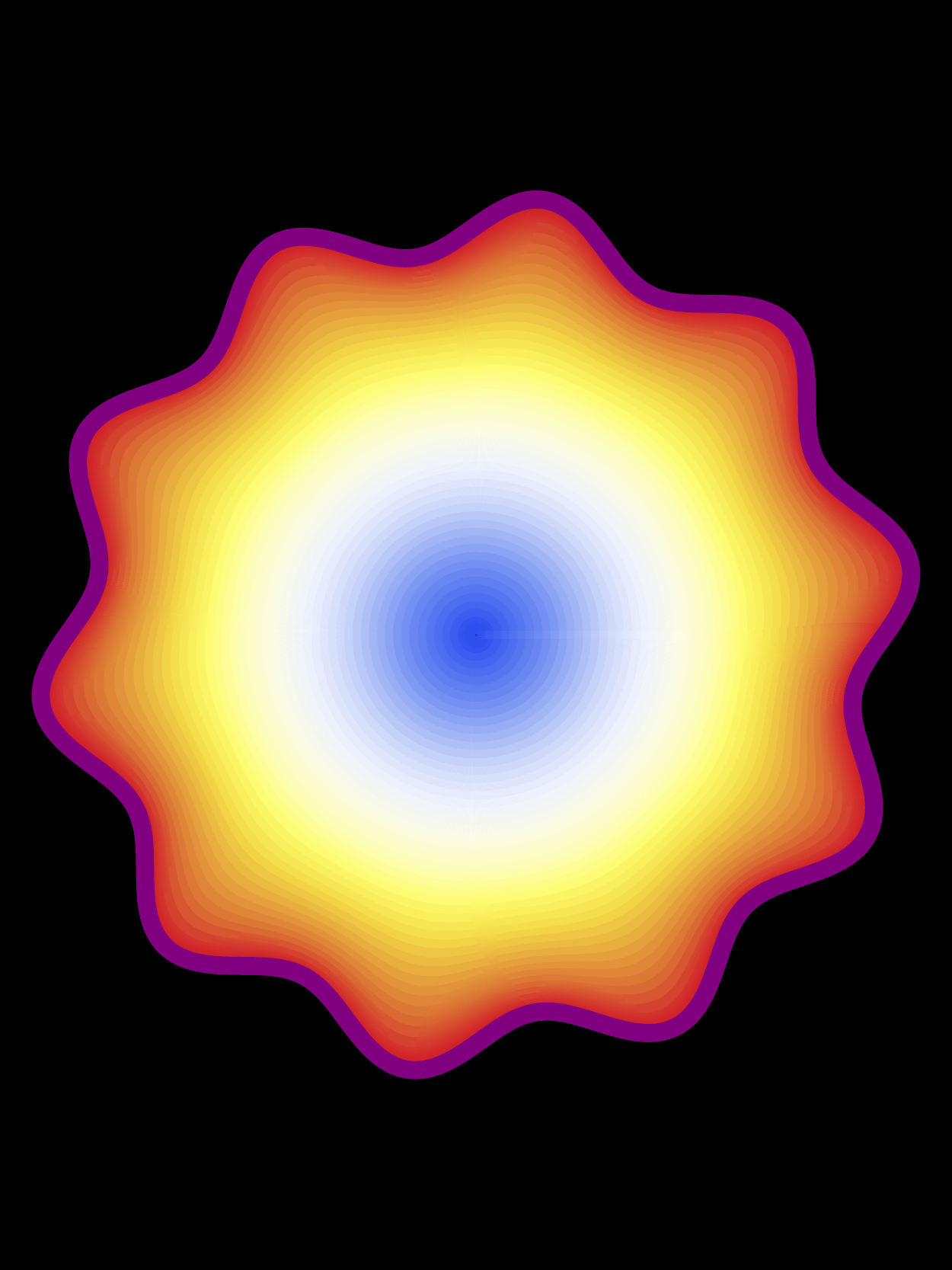}};
        \node[inner sep=1pt,fill=white,draw,rounded corners=1pt,below right=2mm and 2mm] at (figa2.north west) {(b)};
    \end{tikzpicture}
\caption{(a) A pumpkin (white swan acorn variety) with a diameter of 20 cm, featuring circumferential waviness resulting from the drying of its inner pulp during maturation. (b) A bifurcation analysis of a coated elastic disc provides some insight in the complicated problem of pumpkin ripening. In particular, the incremental displacement field (where colours denote intensity increasing fromthe centre to the boundary) of a coated circular disc at bifurcation, involving amodewith 10 wavelengths (with a measured ratio between the disc’s radius and the coating’s thickness of $7.75$, and a Poisson’s ratio of $0.5$), enables the evaluation of the ratio between the Young’s moduli of the pulp and that of the skin. This ratio is found to be equal to $0.6$, a value that is not easily determinable otherwise. }
\label{zucca}
\end{figure}

Coatings are widely used in various technologies, making the findings of this article applicable
to several areas. For instance, the results may be useful in the design of mechanical rollers, as
well as of coated fibres, at both the micro and nanoscales. Another exciting application is to the morphogenesis and growth of plants and fruits. In such cases, turgor pressure can reach up to 10 atmospheres, which is sufficient to trigger various types of bifurcation. For example, it can produce ruffle-like or dome-like shapes in leaves \cite{boudaoud2010introduction} and undulations in flowers, characterized by annular geometry \cite{green1996phyllotactic}. These findings suggest that coatings may play a significant role
in shaping and enhancing the growth of plants and fruits, which can have implications for
agriculture. In fact, the coated disc analysed here exhibits bifurcation modes that result in elegant shapes, resembling those observed during the maturation of some fruits or vegetables. In these cases, a soft pulp is enclosed in a stiff husk, so that pressure generated during drying may lead to the formation of gracious undulations. A prime example is shown in figure \ref{zucca}$a$, which illustrates a pumpkin (white swan acorn variety) and its cross section. 
At the initial growth stage, the
fruit appears smooth on the outside. However, during maturation, the inner pulp dries up and generates a state of compression in the stiff husk, causing it to buckle and resulting in a wavy surface. 
Drying phenomena are highly complex as the case of colloidal drops shows \cite{boulogne2013buckling,bouchaudy2019drying}, nevertheless our results can provide some insight in the maturation problem of the pumpkin. In particular, the bifurcation of a coated disc reveals that the presence of an inner core leads
to complex undulated bifurcation modes, rather than the simple ovalization that occurs in the absence of any inner reinforcement. Figure \ref{zucca}$b$ showcases the outcome of our bifurcation analysis for a coated elastic disc, illustrating the incremental displacement field (colours evidence growing values from the centre to the boundary) for a mode characterized by 10 wavelengths, generated by a coated elastic disc (with a ratio $R/h=7.75$ evaluated for the pumpkin, where $R$ is the radius of the disc and $h$ is the thickness of the circular rod). 
By selecting for the pulp of pumpkin a Poisson’s ratio equal to 0.5 and imposing the observed waviness of the skin, the bifurcation analysis permits the evaluation of the ratio between Young’s moduli of the skin and of the pulp, which is found to fall between $0.543$ and $0.736$ (the figure is generated at $0.6$). Therefore, while a direct measure
may not be easy, our closed-form solution allows to determine the stiffness ratio between husk and pulp by simply counting the number of external undulations without cutting the vegetable.

\section{Large deformation of a planar elastic rod}

\subsection{Kinematics of a curved rod}

\begin{figure}[ht!]
\centering
\begin{tikzpicture}
		\node[inner sep=0pt] (figa2) at (0,0)
{\includegraphics[keepaspectratio, scale=0.9]{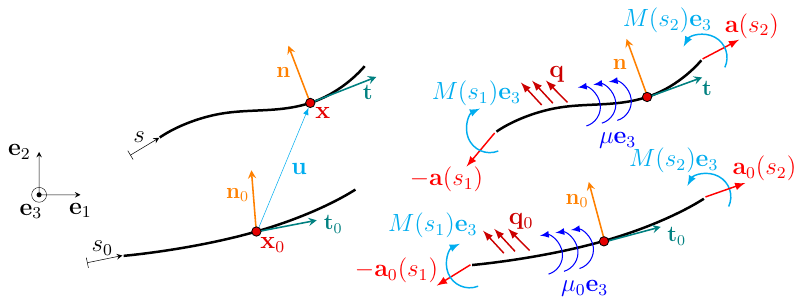}};
		\node[inner sep=1pt,fill=white,draw=none,rounded corners=1pt,below right=2mm and 1mm] at (figa2.north west) {(a)};
    \node[inner sep=1pt,fill=white,draw=none,rounded corners=1pt,below right=2mm and -6 cm] at (figa2.north east) {(b)};
		\end{tikzpicture}
\caption{(a) The deformation $\mathbf{g}\left(\mathbf{x}_{0}\right)$ maps a rod from its reference configuration (characterized by points $\mathbf{x}_0$, arclength parameter $s_0$, unit tangent $\mathbf{t}_0$ and normal $\mathbf{n}_0$) to the deformed configuration (characterised by points $\mathbf{x}$, arclength parameter $s$, unit tangent $\mathbf{t}$ and normal $\mathbf{n}$) through the displacement $\mathbf{u}(\mathbf{x}_0)$. (b): The external and internal forces acting on the current configuration and their counterparts in the reference configuration.
}
\label{kinematics}
\end{figure}

Figure \ref{kinematics}$a$
shows a rod that has undergone deformation in a plane defined by two unit vectors, $\mathbf{e}_1$ and $\mathbf{e}_2$. The reference configuration of the rod is characterized by the arclength parameter $s_0$, while its current configuration is characterized by the arclength parameter $s$. 
The points on the rod in the reference and current configurations are parametrized, respectively, as $\mathbf{x}_0\left(s_0\right)$ and $\mathbf{x}\left(s\right)$ and they are related to each other through the deformation $\mathbf{g}(\mathbf{x}_0)$ and its inverse $\mathbf{g}^{-1}(\mathbf{x})$,
\begin{equation}
    \mathbf{x}(s_0) = \mathbf{g}(\mathbf{x}_0(s_0)) , \quad \mathbf{x}_0(s) = \mathbf{g}^{-1}(\mathbf{x}(s)) .
\end{equation}
The displacement of a point on the rod is defined as 
\begin{equation}
    \mathbf{u} = \mathbf{x} - \mathbf{x}_0 ,
\end{equation}
which can be expressed as a function of either $s_0$ or $s$. 

The unit tangents, $\mathbf{t}_0$ and $\mathbf{t}$, principal normals, $\mathbf{n}_0$ and $\mathbf{n}$, and curvatures, $\kappa_0$ and $\kappa$, are defined in the reference and current configurations, respectively, by
\begin{equation}
    \label{tangent}
    \begin{array}{lll}
        \displaystyle 
        \mathbf{t}_0 = \frac{\partial \mathbf{x}_0}{\partial s_0},  
        &
        \displaystyle 
        \mathbf{t} = 
        \frac{\partial \mathbf{x}}{\partial s}, \\[5mm]
        \displaystyle 
        \mathbf{n}_0 =  \frac{1}{\kappa_0} \frac{\partial \mathbf{t}_0}{\partial s_0}, 
        \quad
        \kappa_0=|\mathbf{t}'_0|,
        &
        \displaystyle 
        \mathbf{n} = \frac{1}{\kappa} \frac{\partial \mathbf{t}}{\partial s}, 
        \quad
        \kappa=|\mathbf{t}'|. 
    \end{array}
\end{equation}

When axial deformation of the rod is negligible and thus axial inextensibility is enforced, the stretch $\lambda$, defined as the ratio between the strained and referential elements $ds$ and $ds_0$, becomes unity, $ds = ds_0$. In this case, the geometrical elements (\ref{tangent}) are related to each other through the equations
\begin{equation}
\label{tt0}
    \mathbf{t} = 
    \frac{\partial \mathbf{u}}{\partial s} + \mathbf{t}_0 , 
    \quad
    \kappa\, \mathbf{n} = \frac{\partial^2\mathbf{u}}{\partial{s}^2} + \kappa_0\mathbf{n}_0 .
\end{equation}

It is instrumental to define unit normals to the rods in the two configurations, $\mathbf{m}_0$ and $\mathbf{m}$, which
coincide with the principal normals, except possibly for the sign, 
\begin{equation}
\label{ms}
    \mathbf{m}_0 = \mathbf{t}_0 \times \mathbf{e}_3 ,
    \quad
    \mathbf{m} = \mathbf{t} \times \mathbf{e}_3 ,
\end{equation}
where $\mathbf{e}_3 = \mathbf{e}_1 \times \mathbf{e}_2$ is the out-of-plane unit vector. The derivative of equations (\ref{ms})with respect to the arclengths $s_0$ and $s$, respectively, leads to 
\begin{equation}
\label{mprime}
    \frac{\partial \mathbf{m}_0}{\partial s_0} = \kappa_0 \mathbf{n}_0 \times \mathbf{e}_3 ,
    \quad
    \frac{\partial \mathbf{m}}{\partial s} = \kappa \mathbf{n}  \times \mathbf{e}_3 ,
\end{equation}
which can alternatively be expressed as 
\begin{equation}
    \frac{\partial \mathbf{m}_0}{\partial s_0} =  - \mbox{sgn}(\mathbf{n}_0 \cdot \mathbf{m}_0)\, \kappa_0\, \mathbf{t}_0 ,
    \quad
    \frac{\partial \mathbf{m}}{\partial s}  = - \mbox{sgn}(\mathbf{n} \cdot \mathbf{m})\, \kappa\, \mathbf{t} .
\end{equation}

The unit vector $\mathbf{m}$ in equation \eqref{ms} and its derivative can be written in terms of referential quantities as 
\begin{equation}
\label{mB0}
    \mathbf{m}=
    \frac{\partial{\mathbf{u}}}{\partial{s}} \times {\mathbf{e}_3}
     + \mathbf{m}_0 , 
     \quad
     \frac{\partial \mathbf{m}}{\partial s} = \left(\frac{\partial^2{\mathbf{u}}}{\partial{s}^2} + \kappa_0\mathbf{n}_0\right)\times \mathbf{e}_3 .
\end{equation}

\subsection{Statics of a curved rod \label{finiteSec}}

\subsubsection{Equilibrium in  the current configuration \label{StatSec}}

An element of the rod in its current configuration, ideally \lq excised' between the arclengths $s_{1}$ and $s_{2}$, is subject to a distributed load $\mathbf{q}(s)$ and moment $\mu(s) \mathbf{e}_{3}$. To mantain equilibrium, internal forces $\mathbf{a}(s)$ and bending moment $M(s)\mathbf{e}_{3}$ must act at the ends $s_{1}$ and $s_{2}$, as shown in figure \ref{kinematics}$b$. The translational equilibrium of the rod is expressed as
\begin{equation}
\label{equilibrium1}
    \mathbf{a}(s_{2}) - \mathbf{a}(s_{1}) + \int_{s_{1}}^{s_{2}}{\mathbf{q}\, ds} = \mathbf{0} ,
\end{equation}  
while the rotational equilibrium is  
\begin{equation}
\label{equilibrium2}
    \big[\,M\mathbf{e}_{3} + \left(\mathbf{x}-\mathbf{o}\right)\times{\mathbf{a}}\big]_{s_{1}}^{s_{2}} + 
    \mathbf{e}_{3} \int_{s_{1}}^{s_{2}}{\mu\, ds} + 
    \int_{s_{1}}^{s_{2}}{\left(\mathbf{x}-\mathbf{o}\right)\times{\mathbf{q}}\, ds} = \mathbf{0} ,
\end{equation}
where $\mathbf{x}-\mathbf{o}$ represents the position vector of a generic point $\mathbf{x}$ on the rod. 

Equations \eqref{equilibrium1} and \eqref{equilibrium2} can be reduced to a unique integral, which can eventually be localized to yield 
\begin{equation}
\label{equilibriumT}
    \frac{\partial{\mathbf{a}}}{\partial{s}} = -\mathbf{q} ,
    \quad
    \frac{\partial{M}}{\partial{s}} + \mu - \mathbf{m}\cdot{\mathbf{a}} = 0 .
\end{equation}

The internal force $\mathbf{a}$ can be defined in terms of axial and shear components, $N$ and $T$, both referred to the current configuration, as 
\begin{equation}
\label{as}
    \mathbf{a} = N\, \mathbf{t} + T\, \mathbf{m} .
\end{equation}
Substitution of equation \eqref{as} into equations (\ref{equilibriumT}) leads to the equilibrium equations,
\begin{equation}
\label{eqtrScalarTot}
    \begin{aligned}
      & \dfrac{\partial{N}}{\partial{s}} - \sgn(\mathbf{n}\cdot{\mathbf{m}})\, \kappa\, T = -\mathbf{q}\cdot{\mathbf{t}}, \\
      & \sgn(\mathbf{n}\cdot{\mathbf{m}})\, \kappa\, N + \dfrac{\partial{T}}{\partial{s}} = -\mathbf{q}\cdot{\mathbf{m}}, \\
      & \dfrac{\partial{M}}{\partial{s}} = T - \mu, 
    \end{aligned}
\end{equation}
holding for any curved rod.

\subsubsection{Equilibrium in the reference configuration}

The equilibrium equations (\ref{eqtrScalarTot}) are now re-derived in the referential description, by introducing the  nominal, or \lq Piola', internal force $\mathbf{a}_0$, defined in a way that
\begin{equation}
\label{piopiopio}
    \mathbf{a}_0 = \mathbf{a} ,
\end{equation}
and with referential, axial and shear force components 
\begin{equation}
\label{eqr2}
    \begin{aligned}
		\mathbf{a}_{0} = N_{0}\mathbf{t}_{0} + T_{0}\mathbf{m}_{0} .
	\end{aligned}
\end{equation}

Note that the bending moment $M$ remains unchanged in the reference configuration, say, $M_0=M$. The same procedure used in the spatial treatment of the equilibrium leads now to 
\begin{equation}
\label{equil0}
\begin{aligned}
    & \dfrac{\partial{N_0}}{\partial{s}} - \sgn(\mathbf{n}_0\cdot{\mathbf{m}_0})\, \kappa_0\, T_0 = -\mathbf{q}_0 \cdot{\mathbf{t}_0}, \\
    & \sgn(\mathbf{n}_0\cdot{\mathbf{m}_0})\, \kappa_0\, N_0 + \dfrac{\partial{T_0}}{\partial{s}} = -\mathbf{q}_0 \cdot{\mathbf{m}_0}, \\
	  & \frac{\partial{M_{0}}}{\partial{s}} - \mathbf{m} \cdot{\mathbf{a}_{0}} = -\mu_{0},
\end{aligned}
\end{equation}
where the external loads in the reference configuration remain unchanged with respect to the deformed configuration 
\begin{equation}
    \mathbf{q}_0 = \mathbf{q} , \quad \mu_0 = \mu ,
\end{equation}
because of the validity of the inextensibility constraint. 

Equation \eqref{ms} yields 
\begin{equation}
\label{t0n0}
\mathbf{t}_{0}\cdot{\mathbf{m}} = -\mathbf{m}_{0}\cdot{\mathbf{t}} , 
\quad 
\mbox{ and } 
\quad
\mathbf{m}_{0}\cdot{\mathbf{m}} = \mathbf{t}_{0}\cdot{\mathbf{t}} ,
\end{equation}
so that substitution of the expression \eqref{eqr2} into equations (\ref{equil0}) leads to the equilibrium equations for a curved rod in the referential description
\begin{equation}
\label{eq120}
\begin{aligned}
    & \dfrac{\partial{N_0}}{\partial{s}} - \kappa_0\, \sgn(\mathbf{n}_0\cdot{\mathbf{m}_0})T_0 + \mathbf{q}\cdot{\mathbf{t}_0} = 0 , \\
	& \kappa_0\, \sgn(\mathbf{n}_0\cdot{\mathbf{m}_0})N_0 + \dfrac{\partial{T_0}}{\partial{s}} + \mathbf{q}\cdot{\mathbf{m}_0} = 0 , \\
    & \dfrac{\partial{M_{0}}}{\partial{s}} =  T_{0}\left(\dfrac{\partial{\mathbf{u}}}{\partial{s}}\cdot{\mathbf{t}_{0}+1}\right) - N_{0}\dfrac{\partial{\mathbf{u}}}{\partial{s}}\cdot{\mathbf{m}_{0} - \mu_{0}} .
\end{aligned}
\end{equation}
It should be noted that equation (\ref{eq120})$_{3}$ can be rewritten as 
\begin{equation}
\label{rotAlt}
    \dfrac{\partial M_0}{\partial s } = 
    T_{0}\, \mathbf{t} \cdot{\mathbf{t}_{0}} - 
    N_{0}\, \mathbf{t} \cdot{\mathbf{m}_{0}} - \mu_{0} .
\end{equation}

\subsubsection{Constitutive equations}

Constitutive equations cannot determine the rod’s axial and shear forces, which are to be understood as reactions to the inextensibility and unshearability constraints. However, the bending moment, which is independent of the referential and spatial distinctions, $M=M_0$, is determined by the difference in the curvature
\begin{equation}
\label{costit}
    M = B\, \frac{\partial \left( \omega - \omega_0 \right)}{\partial s} ,
\end{equation}
where $B=EJ$ is the bending stiffness, equal to the product between Young’s modulus, $E$, of the rod and the second moment of inertia of its cross section, $J$, and $\omega$ and $\omega_0$ are the angles between the tangent vectors in the current and reference configurations and the axis $\mathbf{e}_1$, so that $\mathbf{t} \cdot \mathbf{e}_1 = \cos\omega$ and $\mathbf{t}_0 \cdot \mathbf{e}_1 = \cos\omega_0$. 

\subsection{Incremental equations for a curved rod}
The incremental form of the equilibrium equations for the elastic rod can directly be obtained
from equations \eqref{eqtrScalarTot}, holding for finite deformation, as
\begin{equation}
\label{incrEquilibriumB}
\begin{aligned}
    & \dfrac{\partial{\dot{N}}}{\partial{s}} - \sgn(\mathbf{n}\cdot\mathbf{m})\left( \dot{\kappa}\, T + \kappa\, \dot{T} \right) = -\dot{\mathbf{q}} \cdot{\mathbf{t}} - \mathbf{q}\cdot{\dot{\mathbf{t}}} , \\
    & \sgn(\mathbf{n} \cdot\mathbf{m}) \left(\dot{\kappa}\, N + \kappa\, \dot{N}\right) + \dfrac{\partial{\dot{T}}}{\partial{s}} = -\dot{\mathbf{q}} \cdot{\mathbf{m}} - \mathbf{q}\cdot{\dot{\mathbf{m}}}, \\
    & \dfrac{\partial{\dot{M}\left(s\right)}}{\partial{s}} = \dot{T} - \dot{\mu}, 
\end{aligned}
\end{equation}
where increments are denoted with a superimposed dot.

Since the increments of referential quantities are null, $\dot{\mathbf{t}}_0 = \dot{\mathbf{m}}_0 = \mathbf{0}$, the incremental form of equations \eqref{tt0} and \eqref{ms} become 
\begin{equation}
\label{dottmR}
    \dot{\mathbf{t}} = \frac{\partial\dot{\mathbf{u}}}{\partial{s}} , 
    \quad
    \dot{\mathbf{m}} = \frac{\partial\dot{\mathbf{u}}}{\partial{s}}\times\mathbf{e}_3 .
\end{equation}
Hence, equations \eqref{incrEquilibriumB} can be rewritten as
\begin{equation}
\label{equilB}
	\begin{aligned}
	    & \dfrac{\partial{\dot{N}}}{\partial{s}} - \sgn(\mathbf{n}\cdot{\mathbf{m}}) \left(\dot{\kappa}\, T + \kappa\, \dot{T}\right) = -\dot{\mathbf{q}}\cdot\left(\dfrac{\partial{\mathbf{u}}}{\partial{s}} + \mathbf{t}_0\right) - \mathbf{q} \cdot{\dfrac{\partial\dot{\mathbf{u}}}{\partial{s}}} , \\
		& \sgn(\mathbf{n}\cdot{\mathbf{m}}) \left(\dot{\kappa}\, N + \kappa\, \dot{N}\right) + \dfrac{\partial{\dot{T}}}{\partial{s}} = -\dot{\mathbf{q}}\cdot{\left(\dfrac{\partial{\mathbf{u}}}{\partial{s}}\times{\mathbf{e}_3+\mathbf{m}_0}\right)} - \mathbf{q} \cdot{\left(\dfrac{\partial\dot{\mathbf{u}}}{\partial{s}}\times\mathbf{e}_3\right)} , \\
		& \dfrac{\partial{\dot{M}}}{\partial{s}} = -\dot{\mu}+\dot{T} ,
	\end{aligned}
\end{equation}
which are equivalent to equations \eqref{incrEquilibriumB}, but expressed in terms of incremental displacement $\dot{\mathbf{u}}$, instead than $\dot{\mathbf{t}}$ and $\dot{\mathbf{m}}$. 

The incremental versions of the equilibrium equations \eqref{eq120} find their counterpart in the reference configuration as \begin{equation}
\label{incrEquilibriumB0}
    \begin{aligned}
        & \dfrac{\partial{\dot{N}_0}}{\partial{s}}-\sgn(\mathbf{n}_0\cdot\mathbf{m}_0)\left(\dot{\kappa}_{0}T_0+\kappa_0\dot{T}_0\right)=-\dot{\mathbf{q}}_0\cdot{\mathbf{t}_0} , \\
        & \sgn(\mathbf{n}_0\cdot\mathbf{m}_0)\left(\dot{\kappa}_{0}N_0+\kappa_0\dot{N}_0\right)+\dfrac{\partial{\dot{T}_0}}{\partial{s}}=-\dot{\mathbf{q}}_0\cdot{\mathbf{m}_0} , \\
        & \dfrac{\partial{\dot{M}_{0}}}{\partial{s}}=\dot{T}_0+{\dfrac{\partial{\mathbf{u}}}{\partial{s}}}\cdot\left(\dot{T}_0\mathbf{t}_0-\dot{N}_0\mathbf{m}_0\right)
		+{\dfrac{\partial{\dot{\mathbf{u}}}}{\partial{s}}}\cdot\left(T_0\mathbf{t}_0-N_0\mathbf{m}_0\right)-\dot{\mu}_0 .
    \end{aligned}
\end{equation}
The incremental version of the rotational equilibrium equation \eqref{incrEquilibriumB0}$_3$ can  be rewritten as 
\begin{equation}\dfrac{\partial{\dot{M}_{0}}}{\partial{s}}=T_0\left(\dot{\mathbf{t}}\cdot\mathbf{t}_0\right)-N_0\left(\dot{\mathbf{t}}\cdot{\mathbf{m}_0}\right)+\left(\dot{T}_0\mathbf{t}_0-\dot{N}_0\mathbf{m}_0\right)\cdot\mathbf{t}-\dot{\mu}_0, 
\end{equation}
while the incremental version of the constitutive equation (\ref{costit}) is
\begin{equation}
\label{dotcostit}
    \dot{M} = B \, 
    \frac{\partial \, \dot{\omega}}{\partial s}.
\end{equation}

\section{The annular rod \label{annular_Sec}}

\subsection{Governing equations for a circular rod}

\begin{figure}[hbt!]
		\centering
        \includegraphics[scale=0.8,keepaspectratio]{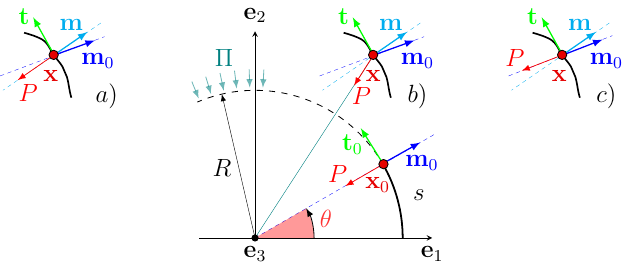}
		\caption{The deformation of an annular rod, where points $\mathbf{x}_0$ of the reference configuration are displaced to points $\mathbf{x}$ of the current configuration. The radial load $\Pi$, uniform in the reference configuration, is assumed to follow three different models under incremental deformation: (a) \textit{hydrostatic pressure}, where the force resultant $P$ over an elementary arc ds follows the normal $\mathbf{m}$ in the deformed configuration, (b) \textit{centrally directed load}, where the resultant $P$ is pointing towards the centre of the rod in the undeformed configuration, (c) \textit{dead load}, where the original direction of the resultant $P$ remains unaltered by deformation. 
  }
		\label{FigAnBeamStr}
	\end{figure}

The theory described above is applicable to rods of any shape, assuming that they are axially inextensible and unshearable, and is particularized now to the case of a circular rod of radius $R$ and centered at point $O$, assumed as origin of a reference system defined by unit vectors $\mathbf{e}_1$ and $\mathbf{e}_2$. The circumferential angle $\theta$ is measured positively in counter-clockwise direction, as depicted in figure \ref{FigAnBeamStr}. 

Due to the polar symmetry being $s=R\theta$, the tangent $\mathbf{t}_0$ and the normal $\mathbf{m}_0$ become
\begin{equation}
\label{t0circ}
    \mathbf{t}_0=-\sin \theta \mathbf{e}_1+\cos \theta \mathbf{e}_2 , 
    \quad 
    \mathbf{m}_0=\cos\theta \mathbf{e}_1+\sin \theta \mathbf{e}_2 ,
\end{equation}
while the position of a generic point $\mathbf{x}_0$ is singled out by $\mathbf{x}_0=R\mathbf{m}_0$. 
Hence, for the annular rod the following relations hold true:
\begin{equation}
\label{relation_Circ}
    \frac{\partial \mathbf{t}_0}{\partial s}=-\frac{1}{R}\mathbf{m}_0, 
    \quad 
    \frac{\partial \mathbf{m}_0}{\partial s}=\frac{\mathbf{t}_0}{R} .
\end{equation}

Equations \eqref{eqtrScalarTot}$_{1-2}$ govern the translational equilibrium of the rod in its spatial configuration
where $\kappa$ represents its deformed curvature. The derivative of the tangent vector $\mathbf{t}$ with respect to the arclength $s$ becomes
\begin{equation}
\label{dtds0Circ}
\begin{aligned}
    \frac{\partial \mathbf{t}}{\partial s}=\frac{\partial^{2}\mathbf{u}}{\partial s^{2}}-\frac{\mathbf{m}_0}{R} ,
\end{aligned}
\end{equation}
so that the curvature is 
\begin{equation}
\label{curvCirc}
    \kappa=\left(\frac{\partial^{2}\mathbf{u}}{\partial s^{2}}-\frac{\mathbf{m}_0}{R}\right)\cdot \mathbf{n},
\end{equation}
to be used in equations \eqref{eqtrScalarTot}$_{1-2}$, which are complemented by 
the rotational equilibrium, equation \eqref{eqtrScalarTot}$_3$.

Equations \eqref{relation_Circ} can be used to express the partial derivative of the internal \lq Piola' force \eqref{eqr2} as 
\begin{equation}
\label{da}
    \frac{\partial \mathbf{a}_0}{\partial s} = \left(\frac{\partial{N_0}}{\partial{s}} + \frac{T_0}{R}\right)\mathbf{t}_0 - \left(\frac{N_0}{R} - \frac{\partial{T_0}}{\partial{s}}\right)\mathbf{m}_0 .
\end{equation}
The displacement vector $\mathbf{u}$ can be expressed in polar components as
\begin{equation}
\label{u0}
    \mathbf{u}=u_{r}\mathbf{m}_{0}+u_{\theta}\mathbf{t}_{0} ,
\end{equation}
so that, according to relations \eqref{relation_Circ}, the derivative with respect to $s$ of the expression \eqref{u0} leads to 
\begin{equation}
\label{u0diff}
\begin{aligned}
    \frac{\partial{\mathbf{u}}}{\partial{s}}=\left(\frac{\partial{u_{r}}}{\partial{s}}-\frac{u_{\theta}}{R}\right)\mathbf{m}_{0}+\left(\frac{u_{r}}{R}+\frac{\partial{u_{\theta}}}{\partial{s}}\right)\mathbf{t}_{0} .
\end{aligned}
\end{equation}
Using equations \eqref{da} and \eqref{u0diff}, the equilibrium equations for the rod in the material configuration are 
\begin{equation}
\label{eq120an}
    \begin{array}{ll}
    \dfrac{\partial{N_{0}}}{\partial{s}}+\dfrac{T_0}{R}=-\mathbf{q} \cdot \mathbf{t}_0,\\[4mm]
    -\dfrac{\partial{T_{0}}}{\partial{s}}+\dfrac{N_0}{R}=\mathbf{q} \cdot \mathbf{m}_0,\\ [4mm]
    \dfrac{\partial{M_{0}}}{\partial{s}}=T_{0}\left(\dfrac{u_{r}}{R}+\dfrac{\partial{u_{\theta}}}{\partial{s}}+1\right)-N_{0}\left(\dfrac{\partial{u_{r}}}{\partial{s}}-\dfrac{u_{\theta}}{R}\right)-\mu_{0} .
    \end{array}
\end{equation}
Kinematic considerations lead to the evaluation of the axial strain $\epsilon$ (to be set equal to zero because of the axial inextensibility), the rotation of cross-section $\Phi$, and the change of curvature $\chi$ as 
\begin{equation}
\label{cinema2}
    \epsilon=\frac{u_{r}}{R} + \frac{\partial u_\theta}{\partial{s}},  
    \quad 
    \Phi = -\frac{u_\theta}{R}+\frac{\partial{u_r}}{\partial{s}},  
    \quad 
    \chi = \kappa-\frac{1}{R} = - \frac{\partial{\Phi}}{\partial{s}} .
\end{equation}
Imposing rod's inextensibility, $\epsilon = 0$, and using equations \eqref{cinema2}, the equation \eqref{eq120an}$_{3}$ can be simplified to  
\begin{equation}
\label{equil}
    \frac{\partial{M_{0}}}{\partial{s}}=T_{0}-\Phi{N_{0}}-\mu_{0} .
\end{equation}
When an external radial load with uniform distribution, $\Pi$, is applied, the annular rod is only subject to a uniform internal compressive force 
\begin{equation}
\label{circP}
	N_{0}=-{\Pi}R , \quad T_{0}=M_{0}=0 ,
\end{equation}
thus, before bifurcation, the rod remains in its circular configuration without suffering any deformation. The incremental quantities \eqref{dottmR} and \eqref{u0diff} assume now the form
\begin{equation}
\label{duincreCirc}
    \dot{\mathbf{t}}=\left(\frac{\partial{\dot{u}_{r}}}{\partial{s}}-\frac{\dot{u}_{\theta}}{R} \right)\mathbf{m}_0 , 
    \quad 
    \dot{\mathbf{m}}=-\left(\frac{\partial{\dot{u}_{r}}}{\partial{s}}-\frac{\dot{u}_{\theta}}{R}\right)\mathbf{t}_{0} , 
    \quad 
    \frac{\partial{\dot{\mathbf{u}}}}{\partial{s}}=\left(\frac{\partial{\dot{u}_{r}}}{\partial{s}}-\frac{\dot{{u}}_{\theta}}{R}\right)\mathbf{m}_{0} ,
\end{equation}
so that the spatial equilibrium is governed by equations \eqref{equilB}, where, from equation \eqref{curvCirc}, the increment in the curvature $\kappa$ is 
\begin{equation}
\label{incrKAppa_Circ}
    \dot{\kappa}=-\frac{\partial^2\dot{\mathbf{u}}}{\partial s^{2}}\cdot\mathbf{m}_0 .
\end{equation}

Taking the material time derivative of the equations \eqref{eq120an}$_{1-2}$, the incremental translational
equilibrium equation in the reference configuration is obtained, so that equation \eqref{circP}$_1$ leads to
\begin{equation}
\label{incrementalEquil_Circ}
    \dfrac{\partial{\dot{N}_{0}}}{\partial{s}}+\dfrac{\dot{T}_{0}}{R}=-\dot{\mathbf{q}}\cdot{\mathbf{t}_{0}} , 
    \quad
    \dfrac{\dot{N}_{0}}{R}-\dfrac{\partial{\dot{T}_{0}}}{\partial{s}}=\dot{\mathbf{q}}\cdot{\mathbf{m}_{0}} ,
    \quad
    \dfrac{\partial{\dot{M}_0}}{\partial{s}}=\dot{T}_{0}+\Pi R\left(\dfrac{\partial{\dot{u}_{r}}}{\partial{s}}-\dfrac{\dot{u}_{\theta}}{R}\right) ,
\end{equation}
thus, from the constitutive equations \eqref{dotcostit} and \eqref{cinema2}$_3$, the left-hand side of equation \eqref{incrementalEquil_Circ}$_{3}$ can be rewritten as
\begin{equation}
\label{momentincr}
	\frac{\partial{\dot{M}_{0}}}{\partial{s}}=-B\left(\frac{\partial^{3}{\dot{u}_{r}}}{\partial{s^{3}}}+\frac{\partial{\dot{u}_{r}}}{\partial{s}}\frac{1}{R^{2}}\right) .
\end{equation}
The use of relation \eqref{momentincr}, when combining together equations \eqref{incrementalEquil_Circ}, yield the differential equations describing the kinematics of the annular rod, subject to an external uniform radial load $\Pi$ 
\begin{equation}
\label{differenzStabilityIncr_Tot}
	 \frac{\partial^{5}{\dot{u}_{r}}}{\partial{\theta^{5}}}+\left(2+\frac{{\Pi}R^{3}}{B}\right)\frac{\partial^{3}{\dot{u}_{r}}}{\partial{\theta^{3}}}+\left(1+2\frac{{\Pi}R^{3}}{B}\right)\frac{\partial{\dot{u}_{r}}}{\partial{\theta}}-\frac{\Pi R^3}{B}\dot{u}_\theta+\mathfrak{
	S}=0 , 
 ~~~
     \dot{u}_{r} + \frac{\partial \dot{u}_\theta}{\partial{\theta}} = 0 ,
\end{equation}
where
\begin{equation}
\label{MStabilityIncr_Tot}
	\mathfrak{S}=-\frac{R^4}{B}\left(\frac{\partial\dot{\mathbf{q}}}{\partial\theta}\cdot{\mathbf{m}}_0+2\dot{\mathbf{q}}\cdot\mathbf{t}_0\right) .
\end{equation}

\subsection{The incremental applied load}

Equation \eqref{differenzStabilityIncr_Tot} requires the evaluation of the increment $\dot{\mathbf{q}}$ in the applied external radial load, taking into account the direction assumed by the load in the deformed configuration \cite{boresi1955refinement,bodner1958conservativeness, boresi1967energy,simitses1968effect, singer1970buckling, batterman1974rigid, schmidt1980critical}.
The modulus $\Pi$ of the radial force is assumed to remain constant, so that $\dot{\Pi}=0$. In particular, as mentioned in the introduction, the three following different cases can be distinguished: 
\begin{itemize}

    \item \textit{Hydrostatic pressure} (i): the applied load remains aligned parallel to the normal $\mathbf{m}$ to the deformed rod element (figure\ref{FigAnBeamStr}$a$), so that the load $\mathbf{q}^{\text{h}}$ and its increment $\dot{\mathbf{q}}^{\text{h}}$ become
        \begin{equation}
        \label{qdotHydro}        
            \mathbf{q}^{\text{h}}=-\Pi\mathbf{m},  
            \quad
         \dot{\mathbf{q}}^{\text{h}}=\Pi\left(\frac{\partial{\dot{u}_{r}}}{\partial{s}}-\frac{\dot{u}_{\theta}}{R}\right)\mathbf{t}_{0} .       
        \end{equation}
   
    \item \textit{Centrally directed load} (ii): the applied load remains directed toward the initial centre of the circular rod (figure \ref{FigAnBeamStr}$b$), so that the load $\mathbf{q}^{\text{l}}$ and its increment $\dot{\mathbf{q}}^{\text{l}}$ become
        \begin{equation}
        \label{qdotCntr}
            \mathbf{q}^{\text{l}}=-\Pi\frac{\mathbf{x}}{\left|\mathbf{x}\right|}, 
            \quad 
            \dot{\mathbf{q}}^{\text{l}}=-\frac{\Pi}{R}\dot{u}_\theta\mathbf{t}_0 .
        \end{equation}
        
     \item \textit{Dead load} (iii): the applied load is dead and does not change its original direction $\mathbf{m}_0$ (figure \ref{FigAnBeamStr}$c$), so that the load $\mathbf{q}^{\text{k}}$ and its increment $\dot{\mathbf{q}}^{\text{k}}$ become
        \begin{equation}
        \label{qdotCns}        
            \mathbf{q}^{\text{k}}=-\Pi\mathbf{m}_0, 
            \quad
            \dot{\mathbf{q}}^{\text{k}}=\mathbf{0} .
        \end{equation}

\end{itemize}
The cases of hydrostatic pressure (i), centrally directed (ii) and dead (iii) load, \cite{singer1970buckling,baazant1991stability,biezeno1945generalized,boresi1967energy,stevens1952stability} are recovered by setting $\mathfrak{S}=\mathfrak{S}^{\Pi}$ in equation \eqref{differenzStabilityIncr_Tot}, being
\begin{equation}
\label{MStabilityIncr_Pressure}
    \mathfrak{S}^{\Pi}=\frac{\Pi R^3}{B}\,\times\left\{
    \begin{array}{lll}
         -\dfrac{\partial\dot{u}_r}{\partial \theta}+\dot{u}_\theta & \text{for hydrostatic pressure (i.)} , \\[3mm]
         \dot{u}_\theta & \text{for centrally directed load (ii.)} , \\[3mm]
         0 & \text{for dead load (iii.)} .
\end{array}\right.
\end{equation} 
Considering a circular elastic rod (without any inner part), the critical loads for bifurcation (i.e. the
smaller buckling load) for the three different loading cases can be derived by imposing continuity of displacement, bending moment and shear force, via equation \eqref{differenzStabilityIncr_Tot}, solved for non-trivial solutions. These critical loads are given by
\begin{equation}
\label{piCr}
    \Pi_{\text{cr}}=
\dfrac{B}{R^3} \, \times 
    \left\{
    \begin{array}{lll}
    3 & \text{for hydrostatic pressure (i)} , \\ 
         9/2 & \text{for centrally directed load (ii)},  \\
         4 & \text{for dead load (iii)} . 
\end{array}\right.
\end{equation} 

Note that the case of hydrostatic pressure does not require any external constraint for equilibrium in the undeformed and deformed configuration. This is different for the other two
loadings (ii) and (iii), where although the undeformed configuration is of equilibrium, the latter is unstable, so that a constraint imposing null translations has to be enforced for case (ii) and a constraint imposing vanishing rotation about the centre has to be enforced for (iii).

\section{Bifurcation of the coated elastic disc \label{AnnulaCoatSect}}
The core of the coated disc is elastic, and for an elastic material under large strain, the constitutive equations, relating the Cauchy stress $\boldsymbol{\sigma}$ to the left Cauchy–Green deformation tensor $\mathbf{B}=\mathbf{F}\mathbf{F}^T$ (where $\mathbf{F}$ is the deformation gradient), can be written as, \cite{gurtin1975continuum}
\begin{equation}
\label{isotrCauchy}
	\boldsymbol{\sigma}=\beta_{0}\mathbf{I}+\beta_{1}\mathbf{B}+\beta_{2}\mathbf{B}^{2} ,
\end{equation}
where the coefficients $\beta_j$ ($j = 0,1,2$) are functions of the  invariants of $\mathbf{B}$. 

The elastic material inside the disc remains unstrained and unstressed up to bifurcation so that $\mathbf{B} = \mathbf{I}$ and $\boldsymbol{\sigma} = \mathbf{0}$. At bifurcation, the incremental relation between the Piola stress $\mathbf{S}$ and the Cauchy stress leads to
\begin{equation}
    \dot{\mathbf{S}} = \dot{\boldsymbol{\sigma}}. 
\end{equation}

The material time derivative of equation \eqref{isotrCauchy} eveals that the elastic response inside the disc follows the usual linear elastic relation, where the Lam\'e constants can be calculated as 
\begin{equation}
    \lambda^{\text{d}} = 2\frac{\partial}{\partial{I_1}}\left(\beta_0+\beta_1+\beta_2\right)+4\frac{\partial}{\partial{I_2}}\left(\beta_0+\beta_1+\beta_2\right) , 
    \quad 
    \mu^{\text{d}} = \beta_1+2\beta_2 ,
\end{equation}
where $I_1$ and $I_2$ are the first and second invariants of the Cauchy-Green deformation tensor, respectively. 

Consider an elastic, homogeneous and isotropic disc characterized by a radius $R$, shear modulus $\mu^{\text{d}}$ and Poisson's ratio $\nu^{\text{d}}$ with reference to a Cartesian coordinate system $(\mathbf{e}_1,\mathbf{e}_2,\mathbf{e}_3)$ with origin $O$ placed at the centre of the disc. A polar reference system ($\mathbf{e}_{r},\mathbf{e}_{\theta},\mathbf{e}_3$) is also introduced, so that the displacement for generalized plane conditions can be written as
\begin{equation}
\label{ud}
    \mathbf{u}^{\text{d}}=u_{r}^{\text{d}}\,\mathbf{e}_{r}+u_{\theta}^{\text{d}}\,\mathbf{e}_{\theta} ,
\end{equation}
where $u^{\text{d}}_{r}$ and $u^{\text{d}}_{\theta}$ are the radial and tangential displacement components. The disc is coated on its boundary by the previously introduced rod with a bending stiffness $B$. In its reference
configuration and loaded with an external radial load $\Pi$, the annular rod is subject to an axial internal force $N_0 = -\Pi R$, while the interior disc remains unstressed.
\begin{figure}[hbt!]
	\centering
    \includegraphics[keepaspectratio, scale=0.8]{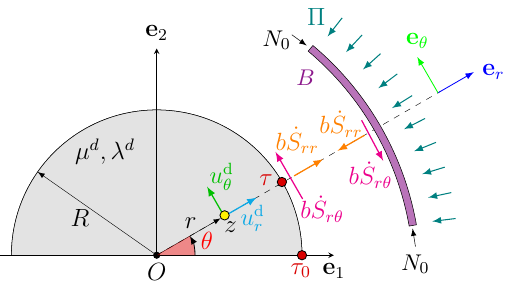}
    \caption{A circular inextensible rod is coating an elastic disc. The bonding between the two is modelled as either perfect or allowing tangential frictionless slip. Loaded by a uniform external radial load, the rod is subject to only an axial force, but at bifurcation incremental internal forces develop so that tractions are transmitted from the external coating to the disc.}
    \label{PressCoatDisk}
\end{figure}
At bifurcation, a non-trivial
incremental deformation occurs, causing the disc to experience incremental stress and strain. The resulting incremental traction at the disc’s boundary, multiplied by its thickness $b$ (to be set equal to the unity for plane strain), gives rise to an incremental force acting on the rod, denoted as $\dot{\mathbf{q}}^{\sigma}$ (figure \ref{PressCoatDisk}). Hence, the incremental load on the coating is given by
\begin{equation}
\label{qdot_PS}
    \dot{\mathbf{q}}=\dot{\mathbf{q}}^{\Pi}+\dot{\mathbf{q}}^{\sigma} ,
\end{equation}
where $\dot{\mathbf{q}}^{\Pi}$ represents the incremental contribution associated with the external radial load $\Pi$ 
\begin{equation}
\label{qsigmadot}
	\dot{\mathbf{q}}^{\sigma}=-b\left(\dot{S}_{rr}\mathbf{m}_0+\mathscr{M}\dot{S}_{r\theta}\mathbf{t}_0\right)_{r=R} ,
\end{equation}
which is determined from the incremental radial and tangential components of the first Piola-Kirchhoff stress tensor $\mathbf{S}$, evaluated on the disc's boundary $r=R$. The term $\mathscr{M}$ in equation \eqref{qsigmadot} describes the shear transmission properties at the interface, so that  $\mathscr{M}=1$ for perfect bonding between disc and coating or $\mathscr{M}=0$  for slip contact, when shear force is not transmissed. As a consequence, equation \eqref{differenzStabilityIncr_Tot} becomes 
\begin{equation}
\label{stability_coat}
\begin{aligned}
	&\frac{\partial^{5}{\dot{u}^{\text{c}}_{r}}}{\partial{\theta^{5}}}+\left(2+\frac{{\Pi}R^{3}}{B}\right)\frac{\partial^{3}{\dot{u}^{\text{c}}_{r}}}{\partial{\theta^{3}}}+\left(1+2\frac{{\Pi}R^{3}}{B}\right)\frac{\partial{\dot{u}^{\text{c}}_{r}}}{\partial{\theta}}-\frac{\Pi R^3}{B}\dot{u}^{\text{c}}_\theta+\mathfrak{S}^{\Pi}+\mathfrak{S}^{\sigma}=0 ,
\end{aligned}
\end{equation}
where the superscript `c' stands for `coating' and
\begin{equation}
\label{mPS}
\begin{aligned}
	\mathfrak{S}^{\text{j}}=-\frac{R^4}{B}\left(\frac{\partial\dot{\mathbf{q}}^{\text{j}}}{\partial\theta}\cdot{\mathbf{m}}_0+2\dot{\mathbf{q}}^{\text{j}}\cdot\mathbf{t}_0\right), && \text{j}=\Pi,\,\sigma .
\end{aligned}
\end{equation}

\subsection{Complex potential formulation for the elastic disc}

In a region enclosed by a sufficiently smooth and non-intersecting curve $L$, each point can be identified with a complex number $z=x_1+ix_2$, where $x_1$ and $x_2$ represent the coordinates of the point and $i$ denotes the imaginary unit. Each point can also be represented in terms of polar coordinates $(r,\theta)$, where $r$ denotes the distance from the origin to the point, and $\theta$ is the angle between $x_1$ and the radius $r$ (measured positively in the counter-clockwise direction), such that $z = re^{i\theta}$. 

The complex displacement $u^{\text{d}}(z)$ in Cartesian coordinates at the point $z$ inside the disc and the complex combination of the normal and shear traction components $\sigma^{\text{d}}(z)$ at that point can be obtained fromthe complex variables forms of the Somigliana’s identities, which are the corollaries of Betti’s reciprocal theorem \cite{linkov1998complex}. The corresponding expressions are
\begin{equation}
\label{somidentity}
\begin{aligned}
    &u^{\text{d}}(z)=\frac{1}{2\pi i(1+\kappa^{\text{d}})}\int_{L}\left[(\kappa^{\text{d}}-1)\frac{u^{\text{d}}(\tau)}{\tau-z}d\tau +u^{\text{d}}(\tau) dK_1(\tau,z)+\overline{u^{\text{d}}(\tau)}dK_2(\tau,z) \right.\\ 
    &\qquad\left.-\frac{\kappa^{\text{d}}}{\mu^{\text{d}}}\sigma^{\text{d}}(\tau)\ln(\tau-z)d\tau+\frac{\kappa^{\text{d}}}{2\mu^{\text{d}}}\sigma^{\text{d}}(\tau)K_1(\tau,z)d\tau-\frac{1}{2\mu^{\text{d}}}\overline{\sigma^{\text{d}}(\tau)}K_2(\tau,z)d\overline{\tau}\right], \\
    &\sigma^{\text{d}}(z)=\frac{\mu^{\text{d}}}{\pi i(1+\kappa^{\text{d}})}\int_{L}\left[ \frac{2\mu^{\text{d}}}{(\tau-z)^2}d\tau-u^{\text{d}}(\tau)\frac{\partial}{\partial z}dK_1(\tau,z)-\overline{u^{\text{d}}(\tau)}\frac{\partial}{\partial z}dK_2+\frac{1-\kappa^{\text{d}}}{2\mu^{\text{d}}}\frac{\sigma^{\text{d}}(\tau)}{\tau-z}d\tau \right. \\
    &\qquad\left. -\frac{\kappa^{\text{d}}}{2\mu^{\text{d}}}\sigma^{\text{d}}(\tau)\frac{\partial K_1}{\partial z}d\tau+\frac{1}{2\mu^{\text{d}}}\overline{\sigma^{\text{d}}(\tau)}\frac{\partial K_2}{\partial z}d\overline{\tau}
    \right],
\end{aligned}
\end{equation}
where a bar over a symbol denotes complex conjugation, $\tau=Re^{i\theta}\in L$, and $u^{\text{d}}(\tau)=u_{1}^{\text{d}}+iu_{2}^{\text{d}}$, $\sigma^{\text{d}}(\tau)=\sigma_{rr}+i\sigma_{r\theta}$ are the displacements and tractions at the boundary, respectively. The kernels $K_1(\tau,z)$, $K_2(\tau,z)$ and the Kolosov constant $\kappa^{\text{d}}$ in equations \eqref{somidentity} are defined as
\begin{equation}
\label{kernKol}
    K_1(\tau,z)=\ln\left(\frac{\tau-z}{\overline{\tau}-\overline{z}}\right), \quad K_2(\tau,z)=\frac{\tau-z}{\overline{\tau}-\overline{z}}, \quad \kappa^{\text{d}} = 
    \left\{
    \begin{aligned}
    & \displaystyle 3 - 4 \nu^{\text{d}}, & \text{for plane strain,} \\
    & \displaystyle \frac{3-\nu^{\text{d}}}{1+\nu^{\text{d}}}, & \text{for plane stress.}
    \end{aligned}
    \right.
\end{equation}
Elastic displacement and stress fields can be determined everywhere in the disc via Kolosov-Muskhelishvili complex potentials $\varphi(z)$ and $\psi(z)$ as  \cite{muskhelishvili2013some} 
\begin{equation}
\label{muskhelishvilifields}
    \left\{
    \begin{aligned}
    & 2\mu^{\text{d}}{u^{\text{d}}\left(z\right)}=\kappa^{\text{d}}\varphi\left(z\right)-z\overline{\varphi^{\prime}\left(z\right)}-\overline{\psi\left(z\right)} , \\
	& \sigma_{11}+\sigma_{22}=4\,\mathrm{Re}\!\left(\varphi^{\prime}\left(z\right)\right) , \\
    & \sigma_{22}-\sigma_{11}+2i\sigma_{12}=2\left[\overline{z}\varphi^{\prime\prime}\left(z\right)+\psi^{\prime}\left(z\right)\right] ,
    \end{aligned} 
    \right.
\end{equation}
where $\mathrm{Re}$ and $\mathrm{Im}$ denote real and the imaginary parts, respectively.

Integral expressions for the complex potentials  $\varphi\left(z\right)$ and $\psi\left(z\right)$ were obtained for a circular disc in  \cite{mogilevskaya2008multiple}, by evaluating the integrals in equation \eqref{somidentity} and using equation \eqref{muskhelishvilifields} \cite{linkov1994complex}, to obtain
\begin{equation}
\label{potentialdisc}
\begin{aligned}
    & \varphi\left(z\right)=\frac{2\mu^{\text{d}}}{\kappa^{\text{d}}-1}\,\mathrm{Re}\!\left(A_{1}\right)\,g^{-1}\left(z\right)+\frac{2\mu^{\text{d}}}{\kappa^{d}}\sum_{n=1}^{\infty}{A_{n+1}\,g^{-\left(n+1\right)}\left(z\right)} , \\
    & \psi\left(z\right)=-\frac{2\mu^{\text{d}}}{\kappa^{\text{d}}-1}\,\mathrm{Re}\!\left(A_{1}\right)\,\frac{\overline{z_{c}}}{R}-\frac{2\mu^{\text{d}}}{\kappa^{\text{d}}}\left[\frac{\overline{z_{c}}}{R}+g\left(z\right)\right]\sum_{n=1}^{\infty}{\left(n+1\right)A_{n+1}\,g^{-n}\left(z\right)} \\
    & \phantom{\psi\left(z\right)=}-2\mu^{\text{d}}\sum_{n=2}^{\infty}{\overline{A_{1-n}}\,g^{-\left(n-1\right)}\left(z\right)} ,
\end{aligned}
\end{equation}
where
\begin{equation}
\label{gz}
\begin{aligned}
    &
    g\left(z\right)=\frac{R}{z}=\frac{R}{\left(x_{1}+ix_{2}\right)} , 
    \quad 
    g^{\prime}\left(z\right)=-\frac{1}{R}\,g^{2}\left(z\right) , 
    \quad
    g^{\prime\prime}\left(z\right)=\frac{2}{R^{2}}\,g^{3}\left(z\right) ,
    \\
    &
    \overline{g\left(z\right)}=\frac{R^{2}}{r^{2}}\,g^{-1}\left(z\right) , 
    \quad
    r=\sqrt{x_{1}^{2}+x_{2}^{2}} ,
\end{aligned}
\end{equation}
and $z_{c}$ denotes the centre of the disc, and $A_{\pm{n}}$ are the complex coefficients in the Fourier series expansions for the displacements on the boundary of the disc, as explained below.

\subsection{Complex combinations for elastic fields on the  disc's boundary\label{cmplx_Sec}}

The complex Fourier series representation for the displacement at every point $\tau=Re^{i\theta}$ on the boundary $L$ of the disc is introduced as
\begin{equation}
\label{displacement}
    u^{\text{d}}\left(\tau\right)=\sum_{n=1}^{\infty}{A_{-n}\,g^{n}\left(\tau\right)}+\sum_{n=0}^{\infty}{A_{n}\,g^{-n}\left(\tau\right)} ,
\end{equation}
where $A_{\pm{n}}$ are unknown complex coefficients and functions $g^{\pm{n}}(\tau)$ are defined from equation \eqref{gz} at $r=R$ as
\begin{equation}
\label{gtau}
    g\left(\tau\right)=\frac{R}{\tau} , 
    \quad 
    \overline{g\left(\tau\right)}=\frac{R}{\overline{\tau}}=g^{-1}\left(\tau\right), 
    \quad
    g^{\prime}\left(\tau\right)=-\frac{1}{R}\,g^{2}\left(\tau\right).
\end{equation}
The relation between Cartesian and polar coordinates,
\begin{equation}
\label{transfrules}
    u^{\text{d}}_{r}\left(\tau\right)+i\,u^{\text{d}}_{\theta}\left(\tau\right)=\left[u^{\text{d}}_{1}\left(\tau\right)+i\,u^{\text{d}}_{2}\left(\tau\right)\right]g\left(\tau\right),
\end{equation}
allows for expressing the displacement components at every point $\tau$ in the polar coordinate system $\left(r,\theta\right)$ as 
\begin{equation}
\label{ur}
    u^{\text{d}}_{r}\left(\tau\right)=\frac{1}{2}\left[u^{\text{d}}\left(\tau\right)\,g\left(\tau\right)+\overline{u^{\text{d}}\left(\tau\right)}\,g^{-1}\left(\tau\right)\right] , 
    \quad 
    u^{\text{d}}_{\theta}\left(\tau\right)=\frac{1}{2i}\left[u^{\text{d}}\left(\tau\right)\,g\left(\tau\right)-\overline{u^{\text{d}}\left(\tau\right)}\,g^{-1}\left(\tau\right)\right] , 
\end{equation}
so that the final representations for displacements are obtained from equation \eqref{displacement} as
\begin{equation}
\label{urComp}
\left.
\begin{aligned}
    & 2u^{\text{d}}_{r}\left(\tau\right) \\
    & 2i\,u^{\text{d}}_{\theta}\left(\tau\right)
\end{aligned}
\right\}=
    \sum_{n=1}^{\infty}{A_{-n}\,g^{n+1}\left(\tau\right)}+\sum_{n=0}^{\infty}{A_{n}\,g^{-\left(n-1\right)}\left(\tau\right)}\pm\sum_{n=1}^{\infty}{\overline{A_{-n}}\,g^{-\left(n+1\right)}\left(\tau\right)}\pm\sum_{n=0}^{\infty}{\overline{A_{n}}\,g^{n-1}\left(\tau\right)} ,
\end{equation}
The complex Fourier series representation for the tractions 
at any point $\tau\in{L}$ are introduced as
\begin{equation} 
\label{traction}
    \sigma^{\text{d}}_{rr}\left(\tau\right)+i\,\sigma^{\text{d}}_{r\theta}\left(\tau\right)=\sum_{n=1}^{\infty}{B_{-n}\,g^{n}\left(\tau\right)}+\sum_{n=0}^{\infty}{B_{n}\,g^{-n}\left(\tau\right)} ,
\end{equation}
where $\sigma^{\text{d}}_{rr}$ and $\sigma^{\text{d}}_{r\theta}$ are the radial and tangential components of the tractions at the point $\tau\in{L}$, respectively, and  $B_{\pm{n}}$ are the unknown complex coefficients. The expressions for the tractions components can be obtained by separating the real and imaginary parts in equation \eqref{traction} as
\begin{equation}
\label{tractionReIm}
\begin{aligned}
    \left.
    \begin{aligned}
       & 2\sigma^{\text{d}}_{rr}\left(\tau\right) \\
        & 2i\,\sigma^{\text{d}}_{r\theta}\left(\tau\right)
    \end{aligned}
    \right\}=
    \sum_{n=1}^{\infty}{B_{-n}\,g^{n}\left(\tau\right)}+\sum_{n=0}^{\infty}{B_{n}\,g^{-n}\left(\tau\right)}\pm\sum_{n=1}^{\infty}{\overline{B_{-n}}\,g^{-n}\left(\tau\right)}\pm\sum_{n=0}^{\infty}{\overline{B_{n}}\,g^{n}\left(\tau\right)} . \\
\end{aligned}
\end{equation}
The complex coefficients $A_{\pm{n}}$ and $B_{\pm{n}}$ are interrelated as \cite{zemlyanova2018circular} 
\begin{equation}
\label{ABinterrelation}
\begin{aligned}
    & B_{-1}=0 , 
    && B_{0}=\frac{4\mu^{\text{d}}}{\left(\kappa^{\text{d}}-1\right)R}\,\mathrm{Re}\!\left(A_{1}\right) , \\
    & B_{-n}=\frac{2\mu^{\text{d}}}{R}(n-1)\,A_{1-n} , \text{ for } n\ge{2} , 
    && B_{n}=\frac{2\mu^{\text{d}}}{\kappa^{\text{d}}R}(n+1)\,A_{n+1} , \text{ for } n\ge{1} .
\end{aligned}
\end{equation}
To satisfy the condition of inextensibility of the coating, additional relations for the complex
coefficients $A_{\pm{n}}$ can be obtained by using equation (99)$_2$ in \cite{mogilevskaya2018elastic} and the following relations \cite{gaibotti2022isoDisk}
\begin{equation}
\label{inexsibility_cmplx}
    \mathrm{Re}\left(A_{1}\right)=0, 
    \quad 
    A_{2}=0, 
    \quad 
    A_{n+1}=\frac{n-1}{n+1}\,\overline{A_{1-n}} 
    \quad 
    \text{for $n\neq{0}$ and $n\neq{-1}$} . 
\end{equation}

\subsection{Complex variable formulation for bifurcation \label{coatedCmplx_Sec}}

Expression \eqref{qsigmadot} can be derived with respect to the arclength $s$ to yield
\begin{equation}
\label{}
    \frac{\partial\dot{\mathbf{q}}^{\sigma}}{\partial{s}}=-\left(\mathscr{M}\frac{\partial{\dot{\sigma}_{r\theta}(\tau)}}{\partial{s}} +\frac{1}{R}\dot{\sigma}_{rr}(\tau)  \right)\mathbf{t}_0 + \left( \frac{\mathscr{M}}{R}\dot{\sigma}_{r\theta}(\tau)-\frac{\partial{\dot{\sigma}_{rr}(\tau)}}{\partial{s}}\right)\mathbf{m}_0 ,
\end{equation}
so that the term $\mathfrak{S}^{\sigma}$ in equation \eqref{mPS}  becomes (for j$=\sigma$)
\begin{equation}
\label{msigma_cmplx}
    \mathfrak{S}^{\sigma}(\tau)=\frac{R^{4}b}{B}\left( R\frac{\partial{\dot{\sigma}_{rr}(\tau)}}{\partial{s}}+\mathscr{M}\dot{\sigma}_{r\theta}(\tau) \right) .
\end{equation}
If the coating is perfectly connected to the disc or if sliding can occur, the following conditions (displacement continuity in the former case, partial continuity and vanishing of shear stress in the latter) have to be imposed, 
\begin{equation}
\label{continuity}
    \dot{u}^{\text{c}}_{r}=\dot{u}^{\text{d}}_{r}\big{|}_{r=R} , 
    \quad
\mbox{and}
\quad    
    \underbrace{\dot{u}^{\text{c}}_{\theta}=\dot{u}^{\text{d}}_{\theta}\big{|}_{r=R}}_{\text{perfect bonding}} 
    \quad 
    \mbox{or}
    \quad
    \underbrace{
    \dot{\sigma}_{r\theta}\big{|}_{r=R} = 0}_{\text{slip contact}}
    ,
\end{equation}
and equation \eqref{stability_coat} can be rearranged as
\begin{equation}
\label{stabcrit0_cmplx}
    \frac{\partial^{5}{\dot{u}^{\text{c}}_{r}}}{\partial{\theta^{5}}}+2\frac{\partial^{3}{\dot{u}^{\text{c}}_{r}}}{\partial{\theta^{3}}}+\frac{\partial{\dot{u}^{\text{c}}_{r}}}{\partial{\theta}}+\frac{\Pi R^3}{B}\left(\frac{\partial^{3}{\dot{u}^{\text{c}}_{r}}}{\partial{\theta^{3}}}+2\frac{\partial{\dot{u}^{\text{c}}_{r}}}{\partial{\theta}}-\dot{u}^{\text{c}}_\theta\right)+\mathfrak{S}^{\Pi}+\mathfrak{S}^{\sigma}=0 .
\end{equation}

Note that the term $\dot{u}^c_\theta$ simplifies in equation \eqref{stabcrit0_cmplx} for all cases, except for the combination of slip contact and dead load, in which case, equation \eqref{stabcrit0_cmplx} has to be differentiated with respect to $\theta$
and the inextensibility condition has to be enforced. A governing sixth-order differential equation
is obtained {\it holding for slip contact and dead radial loading:}
\begin{equation} 
    \label{stabcrit0_deadSplip}
    \frac{\partial^{6}{\dot{u}^{\text{c}}_{r }}}{\partial{\theta^{6}}}+2\frac{\partial^{4}{\dot{u}^{\text{c}}_{r}}}{\partial{\theta^{4}}}+\frac{\partial^{2}{\dot{u}^{\text{c}}_{r}}}{\partial{\theta^{2}}}+\frac{\Pi R^3}{B}\left(\frac{\partial^{4}{\dot{u}^{\text{c}}_{r}}}{\partial{\theta^{4}}}+2\frac{\partial^{2}{\dot{u}^{\text{c}}_{r}}}{\partial{\theta^{2}}}+\dot{u}^{\text{c}}_r\right)+\frac{R^{4}b}{B}\,\frac{\partial^{2}\dot{\sigma}_{rr}}{\partial{\theta^2}}=0 .
\end{equation}
In the following, only the cases pertinent to equation (\ref{stabcrit0_cmplx}) will be explicitely derived, while for the sake of brevity analogous treatment of equation (\ref{stabcrit0_deadSplip}) will not be reported.

Using complex variables formalism, the bifurcation equation \eqref{stability_coat} can be rewritten by adopting the Fourier series representation introduced in Section \ref{cmplx_Sec} for the incremental boundary displacement and stress components at a point $\tau\in{L}$. The first three terms on the left-hand side of equation \eqref{stabcrit0_cmplx}have been already derived in \cite{gaibotti2022isoDisk} and  can now be adapted as 
\begin{equation}
\label{lhsdifferenzialona}
    \frac{\partial^5 \dot{u}_r}{\partial\theta^5}+2 \frac{\partial^3 \dot{u}_r}{\partial\theta^3} + \frac{\partial \dot{u}_r}{\partial\theta}=-R^{3}\,\mathrm{Im}\!\left[\left(R^{2}\,\frac{\partial^{5}{\dot{u}}}{\partial{\tau^{5}}}\,g^{-2}+5R\,\frac{\partial^{4}{\dot{u}}}{\partial{\tau^{4}}}\,g^{-1}+3\,\frac{\partial^{3}{\dot{u}}}{\partial{\tau^{3}}}\right)g^{-2}\right] ,
\end{equation}
(where the superscripts \lq $\text{c}$' and \lq $\text{d}$' have been omitted) 
so that, using equations (43) reported in \cite{gaibotti2022isoDisk}, equation \eqref{lhsdifferenzialona} becomes
\begin{multline}
\label{lhsdifferenzialonaexpr}
    \frac{\partial^5 \dot{u}_r}{\partial\theta^5}+2\, \frac{\partial^3 \dot{u}_r}{\partial\theta^3} + \frac{\partial \dot{u}_r}{\partial\theta}=\frac{1}{2i}\left\{\sum_{n=1}^{\infty}n^{2}\left(n+1\right)\left(n+2\right)^{2}\left[A_{-n}\,g^{n+1}-\overline{A_{-n}}\,g^{-\left(n+1\right)}\right]\right. \\
    \left.
    -\sum_{n=3}^{\infty}{n^2\left(n-1\right)\left(n-2\right)^{2}\left[A_{n}\,g^{-\left(n-1\right)}-\overline{A_{n}}\,g^{n-1}\right]}\right\} .
\end{multline}
The last three terms on the left-hand side of equation \eqref{stabcrit0_cmplx} can be rearranged by combining together equations (99)$_{4-5}$ and (99)$_{1}$ reported in \cite{mogilevskaya2018elastic}, leading to 
\begin{equation}
\label{lhs2differenzialona}
    \frac{\partial^{3}{\dot{u}_{r}}}{\partial{\theta^{3}}}+2\frac{\partial{\dot{u}_{r}}}{\partial{\theta}}-\dot{u}_\theta=R\,\mathrm{Im}\!\left(R^2\,\frac{\partial^{3}{\dot{u}}}{\partial{\tau^{3}}}g^{-2}-\frac{\partial{\dot{u}}}{\partial{\tau}}\right) .
\end{equation}
Employing equations (42) reported in \cite{gaibotti2022isoDisk}, equation  \eqref{lhs2differenzialona} can be written as 
\begin{equation}
\label{lhs2differenzialonaexpr}
\begin{aligned}
    \frac{\partial^{3}{\dot{u}_{r}}}{\partial{\theta^{3}}}&+2\,\frac{\partial{\dot{u}_{r}}}{\partial{\theta}}-\dot{u}_\theta=\frac{1}{2i}\left\{\sum_{n=1}^{\infty}n\left(n^2-3n+1\right)\left[A_{n}\,g^{-\left(n-1\right)}\left(\tau\right)-\overline{A_{n}}\,g^{n-1}\right]\right.\\
    &\left.
    -\sum_{n=1}^{\infty}n\left(n^2+3n+1\right)\left[A_{-n}\,g^{n+1}-\overline{A_{-n}}\,g^{-\left(n+1\right)}\right]\right\} .
\end{aligned}
\end{equation}
Using equation (99)$_1$ reported in  \cite{mogilevskaya2018elastic} in equations \eqref{MStabilityIncr_Pressure}, the complex form of $\mathfrak{S}^{\Pi}$ introduced by equation \eqref{stabcrit0_cmplx} becomes 
\begin{equation}
\label{MSCmplx0_Pressure}
    \mathfrak{S}^{\Pi}=\frac{\Pi R^3}{B}\,\times\left\{
    \begin{array}{lll}
        R\,\mathrm{Im}\!\left(\dfrac{\partial{\dot{u}}}{\partial{\tau}}\right) & \text{for hydrostatic pressure (i.)} ,  \\[3mm]
         \mathrm{Im}\left(\dot{u}\right) & \text{for centrally directed load (ii.)} , \\[3mm]
         0 & \text{for dead load (iii.)} .
\end{array}\right.
\end{equation}
Equations (43) derived in \cite{gaibotti2022isoDisk} and the above equations \eqref{displacement}, allow to rewrite equations \eqref{MSCmplx0_Pressure} 
as
\begin{multline}
\label{MSCmplx_Pressure}
    \mathfrak{S}^{\Pi}=\xi\frac{\Pi R^{3}}{2i\,B}\left\{\sum_{n=1}^{\infty}(-n)^{\alpha}\left[A_{-n}\,g^{n+1}-\overline{A_{-n}}\,g^{-\left(n+1\right)}\right]\right. \\
    \left.+\sum_{n=0}^{\infty}n^{\alpha}\left[A_{n}\,g^{-\left(n-1\right)}-\overline{A_{n}}\,g^{n-1}\right]-\left(\alpha-\xi\right)\left[A_0\,g-\overline{A_0}\,g^{-1}\right]\right\} ,
\end{multline}
where (i) $\xi=\alpha=1$ for hydrostatic pressure, (ii) $\xi=1$, $\alpha=0$ for centrally directed load and (iii) $\xi=0$ for dead load. The term $\mathfrak{S}^{\sigma}(\tau)$ in equation \eqref{stabcrit0_cmplx} is expressed through equation \eqref{msigma_cmplx}, so that the relation (96)$_{2}$ reported in \cite{mogilevskaya2018elastic}, namely, $\partial/\partial{s}=ig^{-1}(\tau)\,\partial/\partial{\tau}$, for points $\tau\in{L}$, leads to  
\begin{equation}
\label{ms0_cmplx}
    \mathfrak{S}^{\sigma}=\frac{R^{4}b}{2i\,B}\left\{ \sum_{n=1}^{\infty}{\left(n+\mathscr{M}\right)}\left[B_{-n}\,g^{n}-\overline{B_{-n}}\,g^{-n}\right] - \sum_{n=0}^{\infty}{\left(n-\mathscr{M}\right)}\left[B_{n}\,g^{-n}-\overline{B_{n}}\,g^{n}\right]
    \right\} .
\end{equation}
Therefore, the bifurcation criterion for the coated disc, equation \eqref{stabcrit0_cmplx}, assumes the form 
\begin{multline}
\label{bucklinggeneral0_cmplx}
    \sum_{n=1}^{\infty}n^{2}\left(n+1\right)\left(n+2\right)^{2}\left[A_{-n}\,g^{n+1}\left(\tau\right)-\overline{A_{-n}}\,g^{-\left(n+1\right)}\left(\tau\right)\right] \\
    -\sum_{n=3}^{\infty}{n^2\left(n-1\right)\left(n-2\right)^{2}\left[A_{n}\,g^{-\left(n-1\right)}\left(\tau\right)-\overline{A_{n}}\,g^{n-1}\left(\tau\right)\right]}+\frac{\Pi{R^3}}{B}\left\{\sum_{n=1}^{\infty}n\left(n^2-3n+1\right)\right. \\
    \left.\left[A_{n}\,g^{-\left(n-1\right)}\left(\tau\right)-\overline{A_{n}}\,g^{n-1}\left(\tau\right)\right]
    -\sum_{n=1}^{\infty}{n\left(n^2+3n+1\right)\left[A_{-n}\,g^{n+1}\left(\tau\right)-\overline{A_{-n}}\,g^{-\left(n+1\right)}\left(\tau\right)\right]}\right\} \\
    +\frac{bR^4}{B}\left\{ \sum_{n=1}^{\infty}{\left(n+\mathscr{M}\right)}\left[B_{-n}\,g^{n}(\tau)-\overline{B_{-n}}\,g^{-n}(\tau)\right]- \sum_{n=0}^{\infty}{\left(n-\mathscr{M}\right)}\left[B_{n}\,g^{-n}(\tau)-\overline{B_{n}}\,g^{n}(\tau)\right]
    \right\} \\
    +\mathfrak{S}^{\Pi}=0 .
\end{multline}
The bifurcation load depends on the particular type of applied radial force per unit length, cases (i)–(iii), while the term $\mathfrak{S}^{\Pi}$ is given by equation \eqref{MSCmplx_Pressure}. Taking into account the expressions for coefficients \eqref{ABinterrelation} and collecting terms with the same power $g^{\pm{n}}(\tau)$ in equation \eqref{bucklinggeneral0_cmplx}, lead to the following result. 
\begin{itemize}
    \item Determination of  coefficients $A_1$ and $A_0$
    \begin{equation}
    \label{ImA1_general}
        \Pi\,\mathrm{Im}\!\left(A_1\right)\left(\xi-1\right)=0 ,
        \quad
        \Pi\,A_{0}\,\xi\left(\alpha-\xi\right)=0 .
    \end{equation}
    \item The bifurcation condition, holding for modes of order $n\geq{2}$: 
    \begin{equation}
    \label{gnBuck_general}
        A_{1-n}\Upsilon(\Pi,n)=0,  
    \end{equation}
    where
    \begin{multline}
    \label{psiPi_general}    
    \Upsilon(\Pi,n)=n^2(n^2-1)+b\frac{\mu^{d}}{B\kappa^{\text{d}}}\left[(n+\mathscr{M})\kappa^{\text{d}}+n-\mathscr{M}\right] \\
    -\Pi\frac{R^3}{B}\left\{ n^2-1-\frac{\xi}{2}\left[\frac{\left(1-n\right)^{\alpha}}{n-1}-\left(n+1\right)^{\alpha-1}\right]\right\}.
    \end{multline}
\end{itemize}

When $A_{1-n}=0$ the trivial solution is obtained, otherwise, 
{\it the bifurcation radial load for the coated disc, corresponding to the $n$-th mode, is:} 
\begin{equation}
\label{np_general}
    \frac{\Pi(n)R^3}{B}=\frac{2n^{2}\left(n^2-1\right)+
    \displaystyle  
    2\frac{\mu^{\text{d}}bR^3}{\kappa^{\text{d}}B}\left[\left(n+\mathscr{M}\right)\kappa^{\text{d}}+n-\mathscr{M}\right]}{2\left(n^2-1\right)+
    \displaystyle 
    \xi\left[(1-n)^{\alpha-1}
    +\left(1+n\right)^{\alpha-1}\right]} ,
    \quad
    n \geq 2, 
\end{equation}
where $\mathscr{M}=1$ ($\mathscr{M}=0$) for perfect bonding (for slip contact) at the rod/core interface and $\xi=\alpha = 1$ for hydrostatic pressure, $\xi=1$ and $\alpha = 0$ for centrally directed load and $\xi=\alpha = 0$  for dead load.

It is worth noting that equation (\ref{np_general}) has been obtained for five load and interface
combinations, excluded the sixth case of slip contact plus dead loading, which requires a separate treatment based on equation (\ref{stabcrit0_deadSplip}). This treatment is omitted for brevity but leads again to equation (\ref{np_general}), which is found to hold true in all cases.  
Equation \eqref{np_general} shows that,  when parameter $\mu^d b R^3/B$  tends to zero, the coated disc behaves as a rod subject to the radial load $\Pi$. For a given set of material and geometrical parameters ($E^{\text{c}}$, $E^{\text{d}}$, $\nu^{\text{d}}$, $R$, $h$, $b$) and varying the mode number $n$ in equation \eqref{np_general}, different values for the bifurcation load can be analyzed. 
The critical value corresponds to the integer number $n$ that minimises equation \eqref{np_general}, so that from the expressions \eqref{gnBuck_general} and \eqref{inexsibility_cmplx}, the only non-vanishing coefficients are $A_{1-n_{\text{cr}}}$ and $A_{1+n_{\text{cr}}}$, define the bifurcation mode. For the elastic rod coating the disc, the latter corresponds to the displacement components
\begin{equation}
\label{critdisp_general}
\begin{aligned}
    & u_{r}(\tau,n)=\mathrm{Re}\!\left(\frac{n}{n+1}A_{1-n}\,g^{n}(\tau)+A_{0}\,g(\tau)\right), \\
    & u_{\theta}(\tau,n)=\mathrm{Im}\!\left(\frac{2}{n+1}A_{1-n}\,g^{n}(\tau)+A_{0}\,g(\tau)+A_{1}\right),
\end{aligned}
\end{equation}
where the non-vanishing coefficient $A_{1-n}$ remains arbitrary, while $A_{0}$ and $\mathrm{Im}\!\left(A_{1}\right)$ can be computed by fixing the rigid-body displacement, as shown in \cite{mogilevskaya2008multiple,  mogilevskaya2018elastic}. The stress components on the boundary of the disc are 
\begin{equation}
\label{critSugma_general}
\begin{aligned}
    & \sigma_{rr}(\tau,n)=\frac{2\mu^{\text{d}}}{R\kappa^{\text{d}}}\left(\kappa^{\text{d}}+1\right)\left(n-1\right)\mathrm{Re}\!\left(A_{1-n}\,g^{n}(\tau)\right), \\
    & \sigma_{r\theta}(\tau,n)=\frac{2\mu^{\text{d}}}{R\kappa^{\text{d}}}\left(\kappa^{\text{d}}-1\right)\left(n-1\right)\mathrm{Im}\!\left(A_{1-n}\,g^{n}(\tau)\right).
\end{aligned}
\end{equation}
All displacement and stress fields at every point $z$ within the boundary of the disc may be determined from equations  \eqref{muskhelishvilifields}, where the complex potentials and their derivatives can be obtained from equation \eqref{potentialdisc} as
\begin{equation}
\label{buckPotential_general}
\begin{aligned}
    & \varphi\left(z,n\right)=\frac{2\mu^{\text{d}}}{\kappa^{\text{d}}}\frac{n-1}{n+1}\,\overline{A_{1-n}}\,g^{-\left(n+1\right)}(z), \\
    & \varphi^{\prime}\left(z,n\right)=\frac{2\mu^{\text{d}}}{R\kappa^{\text{d}}}\left(n-1\right)\,\overline{A_{1-n}}\,g^{-n}(z), \\
    & \varphi^{\prime\prime}\left(z,n\right)=\frac{2\mu^{\text{d}}}{R^{2}\kappa^{\text{d}}}n\left(n-1\right)\,\overline{A_{1-n}}\,g^{-n+1}(z). \\
    & \psi(z,n)=-\frac{2\mu^{\text{d}}}{\kappa^{\text{d}}}\left(n+\kappa^{\text{d}}-1\right)\overline{A_{1-n}}\,g^{-\left(n-1\right)}(z), \\
    & \psi^{\prime}(z,n)=-\frac{2\mu^{\text{d}}}{\kappa^{\text{d}}R}\left(n-1\right)\left(n+\kappa^{\text{d}}-1\right)\,\overline{A_{1-n}}\,g^{-\left(n-2\right)}(z) .
\end{aligned}
\end{equation}

The coefficients $A_0$ and $A_1$ are determined by fixing a rigid-body displacement, in particular, equation  \eqref{muskhelishvilifields}$_1$ assumes the form
\begin{equation}
\label{displrigidfixed}
    u\left(z\right)=\frac{1}{2\mu^{\text{d}}}\left[\kappa^{\text{d}}\varphi\left(z\right)-z\overline{\varphi^{\prime}\left(z\right)}-\overline{\psi\left(z\right)}\right]+A_{0}+i\,\frac{z}{r}\,\mathrm{Im}\!\left(A_{1}\right) ,
\end{equation}
so that the condition $A_0=0$ is obtained by imposing the displacement to be zero at point $z=z_c$ in equation \eqref{displrigidfixed}, as in \cite{gaibotti2022isoDisk}. Again, requiring the displacement component $u_\theta$ to be zero at the point $\tau_0=R$, the following expression for $A_1$ is obtained:
\begin{equation}
\label{imA1fix}
    \mathrm{Im}\!\left(A_{1}\right)=-2\xi\,\mathrm{Im}\!\left(\frac{1}{n+1}A_{1-n}\right) ,
\end{equation}
and hence, under the conditions $A_0=0$ and \eqref{imA1fix}, fixing possible rigid-body roto-translations, the displacement field is determined by equations \eqref{critdisp_general} leading to
\begin{equation}
\begin{aligned}
    &u_r(\tau,n)=\mathrm{Re}\left(A_{1-n}\right)\frac{n}{n+1}\,g^{n}(\tau), \quad u_\theta(\tau,n)=\mathrm{Im}\left(A_{1-n}\right)\frac{2}{n+1}\,\left[g^{n}(\tau)+1\right], \quad \xi=\alpha=1, \\
    &u_r(\tau,n)=\mathrm{Re}\left(A_{1-n}\right)\frac{n}{n+1}\,g^{n}(\tau), \quad u_\theta(\tau,n)=\mathrm{Im}\left(A_{1-n}\right)\frac{2}{n+1}\,\left[g^{n}(\tau)+1\right], \quad \xi=1,\, \alpha=0, \\
    &u_r(\tau,n)=\mathrm{Re}\left(A_{1-n}\right)\frac{n}{n+1}\,g^{n}(\tau), \quad u_\theta(\tau,n)=\mathrm{Im}\left(A_{1-n}\right)\frac{2}{n+1}\,g^{n}(\tau), \quad \xi=0 .
    \end{aligned}
\end{equation}
The displacement amplitude is ruled by the non-vanishing coefficient $A_{1-n}$ which remains arbitrary as in a standard Sturm-Liouville bifurcation problem.

\subsection{Bifurcation results}

\begin{figure}[hbt!]
\centering
    \begin{tikzpicture}
	\node[inner sep=0pt] (figa2) at (0,0) {\includegraphics[width=0.75\textwidth]{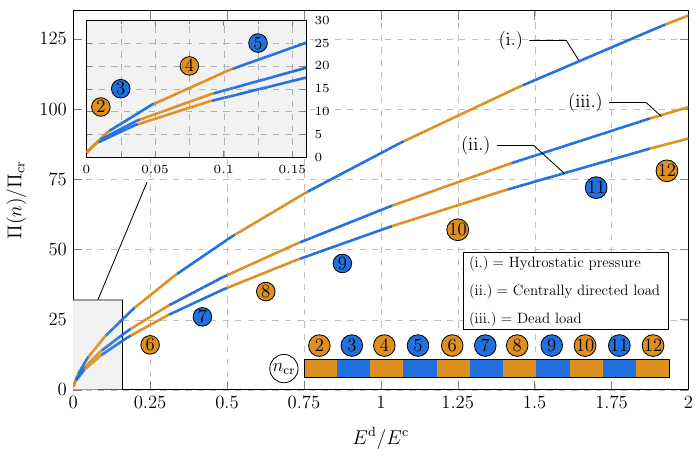}};
	\node[inner sep=1pt,fill=white,draw=none,rounded corners=1pt,below right=2mm and -1 mm] at (figa2.north west) {(a)};
    \end{tikzpicture}
    \begin{tikzpicture}
	\node[inner sep=0pt] (figa2) at (0,0)	{\includegraphics[width=0.75\textwidth]{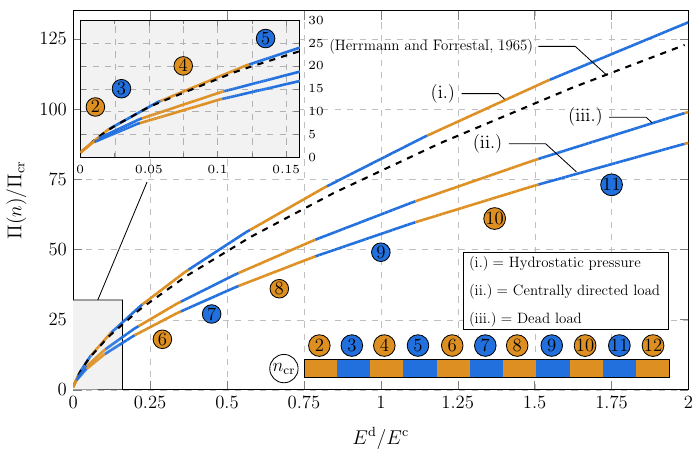}};
	\node[inner sep=1pt,fill=white,draw=none,rounded corners=1pt,below right=2mm and -1mm] at (figa2.north west) {(b)};
    \end{tikzpicture}
    \caption{Dimensionless radial load for bifurcation $\Pi(n)$ as a function of theratio $E^{\text{d}}/E^{\text{c}}$ for hydrostatic ($\xi=\alpha=1$) pressure, centrally directed ($\xi=1$, $\alpha=0$) and dead ($\xi=0$) load. The different colours correspond to the critical bifurcation number $n_{cr}$. (a) Perfect bonding at the disc/coating interface. (b) Slip contact, where the black dashed line corresponds to the approximate solution obtained in \cite{herrmann1965buckling} for a hydrostatic load distribution.
    }
\label{LogTotP}
\end{figure}

The bifurcation radial load, equation \eqref{np_general}, evaluated for the three different types of radial forces per unit length (i)–(iii), as listed in Section \ref{intro}, and for the two models of bonding at the interface, has been normalized through division by the bifurcation load of the ‘empty’ coating, equation  \eqref{piCr}, and has been evaluated as a function of the contrast ratio $E^{\text{d}}/E^{\text{c}}$ between Young’s moduli of the coated disc.
In addition to the latter parameter, the bifurcation load depends on $\kappa^{\text{d}}$ and on the dimensionless parameter $bR^3/[(1+\nu^{\text{d}})\kappa^{\text{d}}J]$. The latter has been assumed equal to 1000, while
the Kolosov constant has been selected as  $\kappa^{\text{d}}=2$, corresponding to $\nu^{\text{d}}=1/4$ in plane strain and $\nu^{\text{d}}=1/3$ in plane stress.
Bifurcation results are reported in figure \ref{LogTotP}, for perfect bonding at the
disc/coating interface (a) and for frictionless slip (b).

Both cases exhibit similar behaviour, in which the critical load increases with the increasing
stiffness contrast between the disc and the coating. However, the critical loads are smaller under
slip conditions than perfect bonding conditions. The increase in the critical load is accompanied by increase in the wavenumber $n$ of the bifurcation mode, evidenced by the alternance of different
colour marking the curves.

\begin{figure}[hbt!]
\centering
    \includegraphics[width=0.75\textwidth]{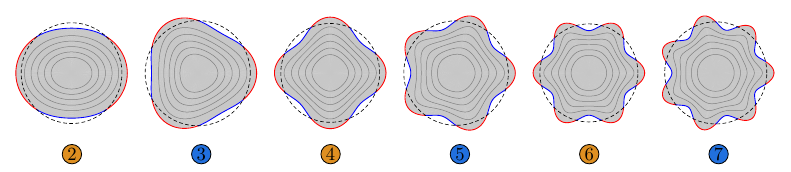}
    \caption{Bifurcation modes for the coated disc at increasing wavenumber $n$. Perfect bonding is assumed at the disc/coating interface. The parts highlighted red (blue) show zones at the interface where tensile (compressive) tractions in the radial direction prevail.}
\label{limoni}
\end{figure}

The shapes of the bifurcation modes are reported in figure \ref{limoni} for the case of perfect bonding,  from $n=2$ to $n=7$. These have been obtained by choosing the non-vanishing coefficient $A_{1-n}=1$ and fixing the rigid body motion accordingly to equation \eqref{imA1fix}. 
The contour of the bifurcation
modes are highlighted red (blue) where tensile (compressive) tractions in the radial direction
prevail. The zones under incremental tension may be expected to detach as a consequence of adhesion failure. 

\begin{figure}[hbt!]
\centering
    \includegraphics[width=0.75\textwidth]{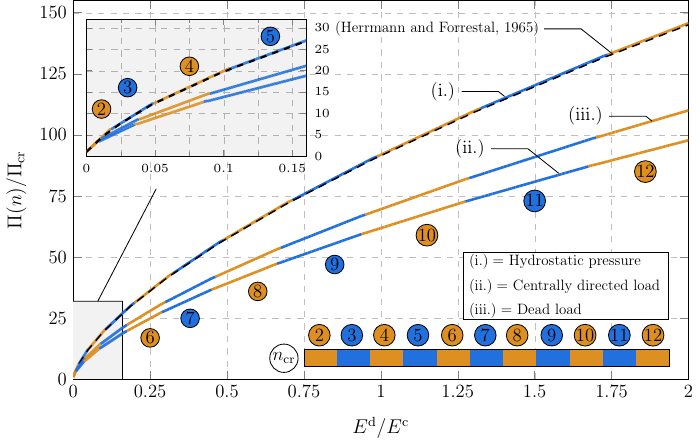}
    \caption{Dimensionless radial load for bifurcation $\Pi(n)$ as a functionof the ratio $E^{\text{d}}/E^{\text{c}}$ for hydrostatic ($\xi=\alpha=1$) pressure, centrally directed ($\xi=1$, $\alpha=0$) and dead ($\xi=0$) load, when slip contact holds at the interface, for $\nu^{\text{d}}=0.455$ under plane strain assumption. The black dashed line corresponds to the approximate solution obtained in \cite{herrmann1965buckling} for a hydrostatic load distribution. This approximate solution matches very well our bifurcation condition for $\nu^{\text{d}}\geq 0.455$.
    }
    \label{Herm0455}
\end{figure}

For the case of slip contact and hydrostatic pressure loading, the results have been compared
with Herrmann and Forrestal \cite{herrmann1965buckling} and included in the figure as a dashed black line. 
The solution by
Herrmann and Forrestal was obtained under the plane strain assumption and by introducing several mathematical approximations, so that for $\nu^{\text{d}}=1/4$ 
the comparison shows the correct
trend, although results are not superimposed. 
Extensive analyses performed by us (and not reported for brevity) show that the approximate solution by Herrmann and Forrestal becomes
tight to our solution $\nu^{\text{d}} \geq 0.455$. In particular, figure \ref{Herm0455} reports the plane strain analysis for $\nu^{\text{d}}=0.455$,0.455, pertaining to the case of slip contact, and shows that the Herrmann and Forrestal approximation is superimposed to our bifurcation curve.

\section{Conclusions}

The Kolosov-Muskhelishvili complex potential technique has proved to enable the analytical
solution of the bifurcation of an elastic disc coated with an inextensible elastic rod. The latter
can be fully bonded or in slip contact with the inner disc and is subject to three different types
of external radial loads, all uniformly distributed. This new solution reveals that the presence
of the inner disc may lead to the predominance of high-wavenumber modes. It also emphasizes
the importance of detachment at the interface between disc and coating, as well as, of proper modelling of how the external load acts on the structure during its incremental deformation.
Applications of the results could be valuable in the development of various coating technologies
and in the understanding of growth of plants and fruits.

\section*{Acknowledgements}
D.B., A.P. and M.G. acknowledge funding from the European Research Council (ERC) under the
European Union’s Horizon 2020 research and innovation programme, grant agreement no. ERC-ADG-2021-101052956-BEYOND. S.G.M acknowledges the support from the National Science Foundation, United States, award no. NSF CMMI - 2112894.

\printbibliography

\end{document}